\newcommand*{\ATLASLATEXPATH}{latex/}
\documentclass[cernpreprint,txfonts,UKenglish,texlive=2011]{\ATLASLATEXPATH atlasdoc}
\pdfoutput=1
\usepackage[subfigure=true]{\ATLASLATEXPATH atlaspackage}
\usepackage{\ATLASLATEXPATH atlasbiblatex}

\usepackage{\ATLASLATEXPATH atlascontribute}

\usepackage[unit=false]{\ATLASLATEXPATH atlasphysics}

\usepackage{slashed}

\usepackage{TOPQ-2014-13-defs}

\newif\ifPrintNotes
\PrintNotestrue
\ifPrintNotes
  \newcommand{\ICBnote}[1]{\textcolor{magenta}{[ICB: #1]}}
  \newcommand{\OAnote}[1]{\textcolor{blue}{[OA: #1]}}
  \newcommand{\DHnote}[1]{\textcolor{green}{[DH: #1]}}
    
\else
  \newcommand{\ICBnote}[1]{\mbox{}}
  \newcommand{\OAnote}[1]{\mbox{}}
  \newcommand{\DHnote}[1]{\mbox{}}
\fi

\newcommand*{\BR}{\ensuremath{\mathcal{B}}\xspace}
\newcommand*{\CL}{\ensuremath{\text{CL}}\xspace}
\newcommand*{\CLs}{\ensuremath{\text{CL}_{s}}\xspace}

\newcommand*{\antikt}{anti-$k_{t}$\xspace}
\newcommand*{\MT}{\ensuremath{m_{\text{T}}}\xspace}
\newcommand*{\MTW}{\ensuremath{m_{\text{T}}(W)}\xspace}
\newcommand*{\mtop}{\ensuremath{m_{\text{top}}}\xspace}
\newcommand*{\WLF}{$W$+LF\xspace}
\newcommand*{\WHF}{$W$+HF\xspace}

\AtlasTitle{Search for single top-quark production via flavour-changing neutral currents at \SI{8}{\TeV} with the ATLAS
detector}

\author{The ATLAS Collaboration}

\AtlasRefCode{TOPQ-2014-13}

\PreprintIdNumber{CERN-PH-EP-2015-172}

\AtlasJournalRef{Eur. Phys. J. C76 (2016) 55}
\AtlasDOI{10.1140/epjc/s10052-016-3876-4}

\AtlasAbstract{%
A search for single top-quark production via flavour-changing neutral current processes
from gluon plus up- or charm-quark initial states in proton--proton collisions at the LHC is presented. 
Data collected with the ATLAS detector in 2012 at a centre-of-mass energy of
\SI{8}{\TeV} and corresponding to an integrated luminosity of \SI{20.3}{\per\fb} are used. 
Candidate events for a top quark decaying into a lepton, a neutrino and a jet are selected and classified 
into signal- and background-like candidates using a neural network. 
No signal is observed and an upper limit on the production cross-section multiplied by the  $t \rightarrow  \ell\nu b$
branching fraction is set. The observed  \SI{95}{\%} \CL limit is 
$\sigma_{qg \rightarrow t} \times \BR (t \rightarrow Wb) \times \BR (W \rightarrow \ell \nu) < \SI{3.4}{\pb}$
and the expected \SI{95}{\%} \CL limit is 
$\sigma_{qg \rightarrow t} \times \BR (t \rightarrow Wb) \times \BR (W \rightarrow \ell \nu) < \SI{2.9}{\pb}$.
The observed limit can be interpreted as upper limits on the coupling constants of the flavour-changing neutral current
interactions divided by the scale of new physics
$\kappa_{ugt}/\Lambda < \SI{10E-3}{\per\TeV}$ and 
$\kappa_{cgt}/\Lambda < \SI{23E-3}{\per\TeV}$ 
and on the branching fractions 
$\BR(t \rightarrow ug) < \num{1.2E-4}$ and 
$\BR(t \rightarrow cg) < \num{6.4E-4}$.
}

\hypersetup{pdftitle={ATLAS paper draft},pdfauthor={The ATLAS Collaboration}}

\begin{document}

\maketitle

\tableofcontents
\clearpage

\section{Introduction}
\label{sec:intro}

The top quark is the most massive elementary particle known, with a mass $\mtop=\SI[parse-numbers=false]{173.3 \pm 0.8}{\GeV}$~\cite{ATLAS:2014wva} close to the electroweak symmetry breaking
scale. 
This makes it an excellent object with which to test the Standard Model (SM) of particle physics,
as well as to search for phenomena beyond the SM. 

At the LHC, top quarks are primarily produced in pairs via the strong interaction. 
In addition to the predominant pair-production process, top quarks are produced singly 
through three different subprocesses via the weak interaction: 
the $t$-channel, which is the dominant process, involving the exchange of a space-like $W$ boson;
the $Wt$ associated production, involving the production of a real $W$ boson;
and the $s$-channel process involving the production of a time-like $W$ boson.

As a consequence of the large value, which is close to one, of the $V_{tb}$ element in the Cabibbo--Kobayashi--Maskawa 
(CKM) matrix, the predominant decay channel of top quarks is $t\to Wb$. 
Transitions between top quarks and other quark flavours mediated by neutral gauge bosons, so-called
flavour-changing neutral currents (FCNC), are forbidden at tree level and suppressed at higher orders
in the SM~\cite{PhysRevD.44.1473}. 
However, several extensions to the SM exist that significantly enhance the production rate 
and hence the branching fractions, \BR, of FCNC processes. Examples of such extensions are the 
quark-singlet model~\cite{AguilarSaavedra:2002ns, delAguila:1998tp, AguilarSaavedra:2002kr}, 
two-Higgs-doublet models with or without flavour conservation~\cite{cheng:1987rs, Grzadkowski:1990sm,
Luke:1993cy, Atwood:1995ud, Atwood:1996vj, Bejar:2000ub}, the minimal supersymmetric standard model~\cite{Li:1993mg, deDivitiis:1997sh, Lopez:1997xv, Guasch:1999jp, Delepine:2004hr, Liu:2004qw, Cao:2007dk} 
or supersymmetry with $R$-parity violation~\cite{Yang:1997dk, Eilam:2001dh}, models with extra 
quarks~\cite{PhysRevD:41891, Arhrib:2006pm, Branco:2013tda}, or the topcolour-assisted technicolour
model~\cite{Lu:2003yr}.
Reviews can be found in Refs.~\cite{AguilarSaavedra:2004wm} and~\cite{Larios:2006pb}.
Many of these models allow for enhanced FCNC production rates, e.g.\ by permitting FCNC interactions
at tree level or introducing new particles in higher-order loop diagrams.
The predicted branching fractions for top quarks decaying to a quark and a neutral boson
can be as large as $10^{-5}$ to $10^{-3}$ 
for certain regions of the parameter space in the models mentioned.
However, the experimental limits have not excluded any specific extension of the SM for the process $t \rightarrow qg$ so far.

\begin{figure}[htbp]
  \centering
  \includegraphics[width=0.45\textwidth]{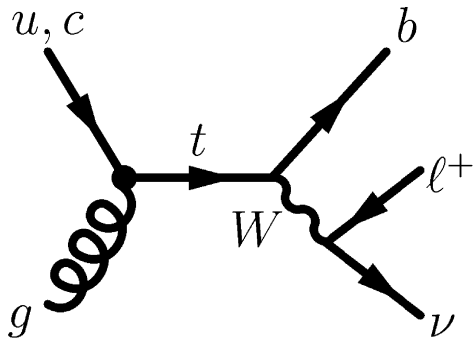}
  \caption{Leading-order Feynman diagram for FCNC top-quark production in the $qg \rightarrow t$ mode followed by the
  decay of the top quark into a $b$-quark and a $W$ boson, where the $W$ boson decays into a lepton and a neutrino.}
  \label{fig:feyndir_pro_decay}
\end{figure}

Among FCNC top-quark decays of the form $t \rightarrow qX$ with $X=Z,H,\gamma,g$, modes involving a $Z$ 
boson, a Higgs boson ($H$), or a photon ($\gamma$) are usually studied directly by searching for final 
states containing the corresponding decay particles.
However, the mode $t \rightarrow qg$, where $q$ denotes either an up quark, $u$, or a
charm quark, $c$, is nearly indistinguishable from the overwhelming background of multi-jet production via
quantum chromodynamic (QCD) processes. 
For the $t \rightarrow qg$ mode, much better sensitivity can be achieved by searching 
for anomalous single top-quark production ($qg\to t$) where a $u$- or $c$-quark and a gluon $g$, originating from the 
colliding protons, interact to produce a single top quark.
A leading-order diagram for top-quark production in the $qg \rightarrow t$ mode as well as a SM decay of the top quark
is shown in Fig.~\ref{fig:feyndir_pro_decay}.%
\footnote{Charge conjugate production and decay modes are implied throughout this paper.}

Anomalous FCNC couplings can be described in a model-independent manner using an effective operator 
formalism~\cite{Buchmueller1986621}, which assumes the SM to be the low-energy limit of a more general theory
that is valid at very high energies. 
The effects of this theory below a lower energy scale, $\Lambda$, are perceived through a set of effective operators
of dimension higher than four. 
The formalism therefore allows the new physics to be described by an effective Lagrangian consisting of the 
SM Lagrangian and a series of higher-dimension operators, which are
suppressed by powers of $1/\Lambda$.
The new physics scale, ${\Lambda}$, has a dimension of energy and is related to the mass cut-off scale 
above which the effective theory breaks down, hence characterising the energy scale at which the new physics manifests
itself in the theory.
A further method for simplifying the formalism is to only consider operators of interest that have no 
sizeable impact on physics below the \si{\TeV} scale, following Ref.~\cite{Ferreira:2006xe}.

The interest of this paper lies in effective dimension-six operators,
which contribute to flavour-changing interactions in the strong sector; thus no operators with electroweak gauge bosons are considered.
In particular, the operators describing FCNC couplings to a single top quark are of interest here;
they describe strong FCNC vertices of the form $qgt$ and can be written as~\cite{Coimbra:2012ys}:
\begin{linenomath}
\begin{equation*}
\mathcal{O}_{uG\Phi}^{\,ij}=\bar{q}_{\mathrm
L}^{\,i}\,\lambda^a\,\sigma^{\mu\nu}\,u_{\mathrm R}^j\,\tilde{\Phi}\,G^{a\mu\nu}\,,
\end{equation*}
\end{linenomath}
where $u_R^j$ stands for a right-handed quark singlet, $\bar{q}_{L}^{\,i}$ for a left-handed quark doublet,
$G^{a\mu\nu} $ is the gluon field strength tensor, $\tilde{\Phi}$ the charge conjugate of the Higgs doublet,
$\lambda^a$ are the Gell-Mann matrices and $\sigma^{\mu\nu}$ is the anti-symmetric tensor.
The indices ($i,j$) of the spinors are flavour indices indicating the quark generation. 
By requiring a single top quark in the interaction, one of the indices can always be set 
equal to 3 while the other index is either 1 or 2.
Hence, the remaining fermion field in the interaction is either a $u$- or a $c$-quark.
Apart from direct single top-quark production, these operators give rise to interactions of the 
form $gg \to tq$ and $gq \to tg$.
The processes considered are a subset of these, where a $u$-quark, $c$-quark or gluon originating
from the colliding protons interacts through an $s$-, $t$- or $u$-channel process to produce a single top quark, 
either via a $(2 \to 2)$ process or without the associated production of additional gluons or 
light quarks via a $(2 \to 1)$ process. 

The corresponding strong FCNC Lagrangian usually is written as~\cite{Coimbra:2012ys}:
\begin{linenomath}
\begin{equation*}
\mathcal{L}_{\mathrm S} = -g_{\mathrm s}
\sum_{q=u,c}\,\frac{\kappa_{qgt}}{\Lambda}\,\bar{q}\,\lambda^a\,\sigma^{\mu\nu}\,(f_{q} + h_{q}\gamma_5)\,t\,G^{a}_{\mu\nu} + \text{h.c.}\,,
\end{equation*}
\end{linenomath}
with the real and positive parameters $\kappa_{gqt}\,(q=u,c)$ that relate the strength of the new couplings to the
strong coupling strength, $g_{\mathrm s}$, and where $t$ denotes the top-quark field.
The parameters $f_{q}$ and $h_{q}$ are real, vector and axial chiral parameters, respectively, which satisfy the relation $|f_{q}|^2 + |h_{q}|^2 = 1$.
This Lagrangian contributes to both the production and decay of top quarks.

Experimental limits on the branching fractions of the FCNC top-quark decay channels have been set
by experiments at the LEP, HERA, Tevatron and LHC accelerators.
At present the most stringent upper 
limits at \SI{95}{\%} confidence level (\CL) for the coupling constants $\kappa_{\gamma qt}$ and $\kappa_{qgt}$ 
are $\kappa_{\gamma qt}/\mtop < \SI{0.12}{\GeV^{-1}}$~\cite{Abramowicz:2011tv} (ZEUS, HERA) and 
$\BR(t\to qg) < \num{5.7E-5}$ ($ugt$) and $\BR(t\to qg) < \num{2.7E-4}$ ($cgt$)~\cite{TOPQ-2011-18} (ATLAS, LHC).
In the case of $t \rightarrow qZ$, upper limits on the branching fractions of the top-quark decay have been 
determined to be $\BR(t\to qZ) < \SI{0.05}{\%}$~\cite{CMS-TOP-12-037} (CMS, LHC).
Finally, the most stringent limit for the decay $t\to qH$ is measured to be $\BR(t\to qH) <
\SI{0.79}{\%}$~\cite{HIGG-2013-09} (ATLAS, LHC).

In the allowed region of parameter space for $\kappa_{qgt}/\Lambda$, the FCNC production 
cross-section for single top quarks is of the order of picobarns, while the
branching fraction for FCNC decays is very small, i.e.\ below \SI{1}{\%}.
Top quarks are therefore reconstructed in the SM decay mode $t\to Wb$.
The $W$ boson can decay into a quark--antiquark pair ($W\rightarrow q_1 \bar{q}_{2}$) or a
charged lepton--neutrino pair ($W\rightarrow\ell\nu$); only the latter is
considered here.
This search targets the signature from the $qg \to t \to W(\to \ell\nu)\,b$ process. 
Events are characterised by an isolated high-energy charged lepton (electron or muon), 
missing transverse momentum from the neutrino and exactly one jet 
produced by the hadronisation of the $b$-quark. 
Events with a $W$ boson decaying into a $\tau$ lepton, where the $\tau$ decays into an
electron or a muon, are also included.
Several SM processes have the same final-state topology and are considered as background to the FCNC analysis. 
The main backgrounds are $V$+jets production (especially in association with heavy quarks), where $V$ denotes a $W$ or a $Z$ boson, 
SM top-quark production, diboson production, and multi-jet production via QCD processes.
The studied process can be differentiated from SM single top-quark production,
which is usually accompanied by additional jets.
Furthermore, FCNC production has kinematic differences from the background processes,
such as lower transverse momenta of the top quark.

This paper is organised as follows: Section~\ref{sec:detector} provides a description of the ATLAS detector.
Section~{\ref{sec:data}} gives an overview of the data and Monte Carlo (MC) samples used for the simulation of
signal and expected background events from SM processes. In \Sect{\ref{sec:evtsel}} the event selection is
presented. The methods of event classification into signal- and background-like events using a neural network are
discussed in \Sect{\ref{sec:analysis}} and sources of systematic uncertainty are summarised
in \Sect{\ref{sec:systematics}}. The results are presented in \Sect{\ref{sec:result}}
and the conclusions are given in \Sect{\ref{sec:conclusion}}.

\section{ATLAS detector}
\label{sec:detector}

The ATLAS detector~\cite{PERF-2007-01} is a multipurpose collider detector built from a set of 
sub-detectors, which cover almost the full solid angle around the 
interaction point.\footnote{ATLAS uses a right-handed coordinate system with its origin at the nominal interaction point in the centre
of the detector and the $z$-axis is along the beam direction;
the $x$-axis points towards the centre of the LHC ring and 
the $y$-axis points upwards. The pseudorapidity 
$\eta$ is defined as $\eta=-\ln[\tan(\theta/2)]$, where the polar angle 
$\theta$ is measured with respect to the $z$-axis. The azimuthal angle, 
$\phi$, is measured with respect to the $x$-axis.
Transverse momentum and energy are defined as $\pT = 
p\sin\theta$ and $\ET = E\sin\theta$, respectively. The $\Delta R$ distance 
in ($\eta$,$\phi$) space is defined as $\Delta R=\sqrt{(\Delta\eta)^2+(\Delta\phi)^2}$.}
It is composed of an inner tracking detector (ID) close to the 
interaction point surrounded by a superconducting solenoid 
providing a \SI{2}{\tesla} axial magnetic field, electromagnetic 
and hadronic calorimeters, and a muon spectrometer (MS).
The ID consists of a silicon pixel detector, a silicon microstrip detector 
providing tracking information within pseudorapidity $|\eta| < 2.5$,
and a straw-tube transition radiation tracker that covers $|\eta| < 2.0$.
The central electromagnetic calorimeter is a lead and liquid-argon (LAr) 
sampling calorimeter with high granularity, and is divided into
a barrel region that covers $|\eta| < 1.475$ and endcap regions 
that cover $1.375 < |\eta| < 3.2$.
An iron/scintillator tile calorimeter
provides hadronic energy measurements in the central pseudorapidity range.
The endcap and forward regions are instrumented with LAr calorimeters for 
both the electromagnetic and hadronic energy
measurements, and extend the coverage to $|\eta | = 4.9$.
The MS covers $|\eta| < 2.7$ and consists of three large superconducting toroids with eight coils each,
a system of trigger chambers, and precision tracking chambers.

\section{Data and simulated samples}
\label{sec:data}
This analysis is performed using $\sqrt{s}= \SI{8}{\TeV} $ proton--proton ($pp$)
collision data recorded by the ATLAS experiment in 2012. 
Stringent detector and data quality requirements are applied, resulting in a
data sample with a total integrated luminosity of \SI{20.3}{\per\fb}.

\subsection{Trigger requirements}
\label{sec:trigger}
ATLAS employs a three-level trigger system for selecting events to be recorded. 
The first level (L1) is built from custom-made hardware, 
while the second and third levels are software based and collectively referred to as 
the high-level trigger (HLT).
The datasets used in this analysis are defined by high-\pT 
single-electron or single-muon triggers~\cite{PERF-2011-02, TRIG-2012-03}.

For the L1 calorimeter trigger, which is based on reduced calorimetric information, a cluster in the electromagnetic calorimeter is
required with $\ET > \SI{30}{\GeV}$ or with $\ET > \SI{18}{\GeV}$. 
The energy deposit must be well separated from other clusters.
At the HLT, the full granularity of the calorimeter and tracking information is available. 
The calorimeter cluster is matched to a track and the trigger electron candidate is required to 
have $\ET > \SI{60}{\GeV}$ or $\ET > \SI{24}{\GeV}$ with additional isolation requirements.

The single-muon trigger is based on muon candidates reconstructed in the MS.
The triggered events require a L1 muon trigger-chamber track 
with a \SI{15}{\GeV} threshold on the $\pT$ of the track. At the HLT, the requirement is
tightened to $\pT > \SI{24}{\GeV}$ with, or \SI{36}{\GeV} without, an isolation criterion.

\subsection{Simulated events}
Simulated event samples are used to evaluate
signal and background efficiencies and uncertainties
as well as to model signal and background shapes.

For the direct production of top quarks via FCNC, \METOP~\cite{Coimbra:2012ys} is used for simulating strong FCNC processes at next-to-leading order (NLO) in QCD. It introduces strong top-quark FCNC interactions through effective operators. 
By comparing kinematic distributions for different FCNC couplings, it has been verified that the kinematics of the 
signal process are independent of the \textit{a priori} unknown FCNC coupling strength.
As a conservative approach, only left-handed top quarks (as in the SM) are produced, and the decay of the top
quark is assumed also to be as in the SM.\footnote{%
A right-handed top quark would give rise to different angular distributions and hence be easier to separate from SM production.}
The \ct10~\cite{Lai:2010vv} parton distribution function~(PDF) sets are used for the 
generation of the signal events and the renormalisation and factorisation scales
are set to the top-quark mass. 

The \textsc{Powheg-box}~\cite{FRI-0701} generator with the \ct10 PDF sets is used to generate 
\ttbar~\cite{FRI-0301} and electroweak single top-quark production in the $t$-channel~\cite{Frederix:2012dh}, 
$s$-channel~\cite{Alioli:2009je} and $Wt$-channel~\cite{Re:2010bp}.
All processes involving top quarks, including the strong FCNC processes, are produced assuming $\mtop = \SI{172.5}{\GeV}$.
The parton shower and the underlying event are added using \PYTHIA~6.426~\cite{Sjostrand:2006za}, where 
the parameters controlling the modelling are set to the values of the Perugia 2011C tune~\cite{Skands:2010ak}.

Vector-boson production in association with jets ($V$+jets) is simulated using the multi-leg leading-order (LO) 
generator \SHERPA~1.4.1~\cite{Gleisberg:2008ta} with its own parameter tune and the \ct10 PDF sets.
\SHERPA is used not only to generate the hard process, but also for the parton shower and
the modelling of the underlying event.
$W$+jets and $Z$+jets events with up to five additional partons are generated.
The CKKW method~\cite{Hoeche:2009rj} is used to remove overlap between 
partonic configurations generated by the matrix element and by parton shower evolution.
Double counting between the inclusive $V$+$n$ parton samples and samples with associated heavy-quark pair production
is avoided consistently by using massive $c$- and $b$-quarks in the shower. 

Diboson events ($WW$, $WZ$ and $ZZ$) are produced using \ALPGEN~2.14~\cite{SAMPLES-ALPGEN} and the \cteq PDF
sets~\cite{cteq6l}.
The partonic events are showered with \HERWIG~6.5.20~\cite{Corcella:2000bw}, and 
the underlying event is simulated with the \JIMMY~4.31~\cite{Butterworth:1996zw} model 
using the ATLAS Underlying Event Tune~2~\cite{ATL-PHYS-PUB-2011-008}.

All the generated samples are passed through the full simulation of the ATLAS 
detector~\cite{SOFT-2010-01} based on \GEANT{}4~\cite{Agostinelli:2002hh} and are then 
reconstructed using the same procedure as for data. The simulation includes the effect of 
multiple $pp$ collisions per bunch crossing. The events are weighted such that 
the average distribution of the number of collisions per bunch crossing is the same as in data.
In addition, scale factors are applied to the simulated events to take into account small differences
observed between the efficiencies for the trigger, lepton identification and $b$-quark jet identification.
These scale factors are determined using control samples.

\section{Event selection}
\label{sec:evtsel}

The expected signature of signal events is used to perform the event selection.
Events containing exactly one isolated electron or muon, 
missing transverse momentum and one jet, which is required to be identified as a jet originating from a $b$-quark, are
selected.

\subsection{Object definition and event selection}
\label{sec:objsel}

Electron candidates are selected from energy deposits (clusters) in the LAr electromagnetic calorimeter 
associated with a well-measured track fulfilling strict quality requirements~\cite{PERF-2010-04}.
Electron candidates are required to satisfy $\pT > \SI{25}{\GeV}$ and 
$|\eta_{\text{clus}}| < 2.47$, where $\eta_{\text{clus}}$ 
denotes the pseudorapidity of the cluster.  Clusters falling in the calorimeter barrel--endcap 
transition region, corresponding to $1.37<|\eta_{\text{clus}}|<1.52$, are ignored.
High-\pT~electrons associated with the $W$-boson decay can be mimicked by hadronic jets reconstructed 
as electrons, electrons from the decay of heavy quarks, and photon conversions.  
Since electrons from the $W$-boson decay are typically isolated from hadronic jet activity, backgrounds can be suppressed by isolation criteria,
which require minimal calorimeter activity 
and only allow low-$\pT$ tracks in an $\eta$--$\phi$ cone around the electron candidate.
Isolation cuts are optimised to achieve a uniform cut efficiency of \SI{90}{\%} as a function of $\eta_{\text{clus}}$ 
and transverse energy, $\ET$.
The direction of the electron candidate is taken as that of the associated track.
For the calorimeter isolation a cone size of $\Delta R = 0.2$ is used. 
In addition, the scalar sum of all track transverse momenta within a cone of size $\Delta R = 0.3$ 
around the electron direction is required to be below a \pT-dependent threshold in the range between \SI{0.9}{\GeV} and \SI{2.5}{\GeV}.
The track belonging to the electron candidate is excluded from this requirement. 

Muon candidates are reconstructed by matching track segments or complete tracks in the MS
with tracks found in the ID~\cite{PERF-2014-05}. The final candidates are required to have a transverse
momentum $\pT > \SI{25}{\GeV}$ and to be in the pseudorapidity region $|\eta|<2.5$.
Isolation criteria are applied to reduce background events in which a high-\pT muon is produced
in the decay of a heavy-flavour quark.
An isolation variable~\cite{Rehermann:2010vq} is defined as the scalar 
sum of the transverse momenta of all tracks with \pT above \SI{1}{\GeV}, except the one matched to the muon,
within a cone of size $\Delta R_{\text{iso}} = \SI{10}{\GeV}/\pT(\mu)$.
Muon candidates are accepted if they have an isolation to $\pT(\mu)$ ratio of less than 0.05. 
An overlap removal is applied between the electrons and the muons,
rejecting the event if the electron and the muon share the same ID track.

Jets are reconstructed using the \antikt algorithm~\cite{Cacciari:2008gp} with a radius parameter of 0.4, 
using topological clusters~\cite{ATL-LARG-PUB-2008-002} as inputs to the jet finding. 
The clusters are calibrated with a local cluster weighting method~\cite{PERF-2012-01}.
Calibrated jets using an energy- and $\eta$-dependent simulation-based
calibration scheme, with \emph{in situ} corrections based on data,
are at first required to have $\pT > \SI{25}{\GeV}$ and $|\eta|<2.5$.
The jet energy is further corrected for the effect of multiple $pp$ interactions, both in 
data and in simulated events.

If any jet is within $\Delta R = 0.2$ of an electron, the closest jet is removed, since in these cases the 
jet and the electron are very likely to correspond to the same physics object. Remaining electron candidates 
overlapping with jets within a distance $\Delta R<0.4$ are subsequently rejected.
To reject jets from pile-up events, a so-called jet-vertex fraction criterion is applied
for jets with $\pT < \SI{50}{\GeV}$ and $|\eta| <2.4$:
at least \SI{50}{\%} of the scalar sum of the \pT of tracks within a jet 
is required to be from tracks compatible with the 
primary vertex\footnote{The primary vertex is defined as the vertex with the largest $\sum \pT^2$ of the associated
tracks.} associated with the hard-scattering collision. 
The final selected jet is required to have $\pT > \SI{30}{\GeV}$ and must also be identified as a
jet originating from a $b$-quark ($b$-tagged).

In this analysis, a $b$-tagging algorithm that is optimised to improve the rejection of $c$-quark jets is used,
since $W+c$ production is a major background.
A neural-network-based algorithm is used, which combines three different algorithms exploiting 
the properties of a $b$-hadron decay in a jet~\cite{ATLAS-CONF-2011-102}.
The chosen working point corresponds to a $b$-tagging efficiency of \SI{50}{\%}, when cutting on the discriminant, and a $c$-quark jet and light-parton jet mistag acceptance of \SI{3.9}{\%} and \SI{0.07}{\%},
respectively, as measured in \ttbar events~\cite{ATLAS-CONF-2014-046,ATLAS-CONF-2014-004}.

\begin{figure}[htbp]
  \centering
  \includegraphics[width=0.7\textwidth]{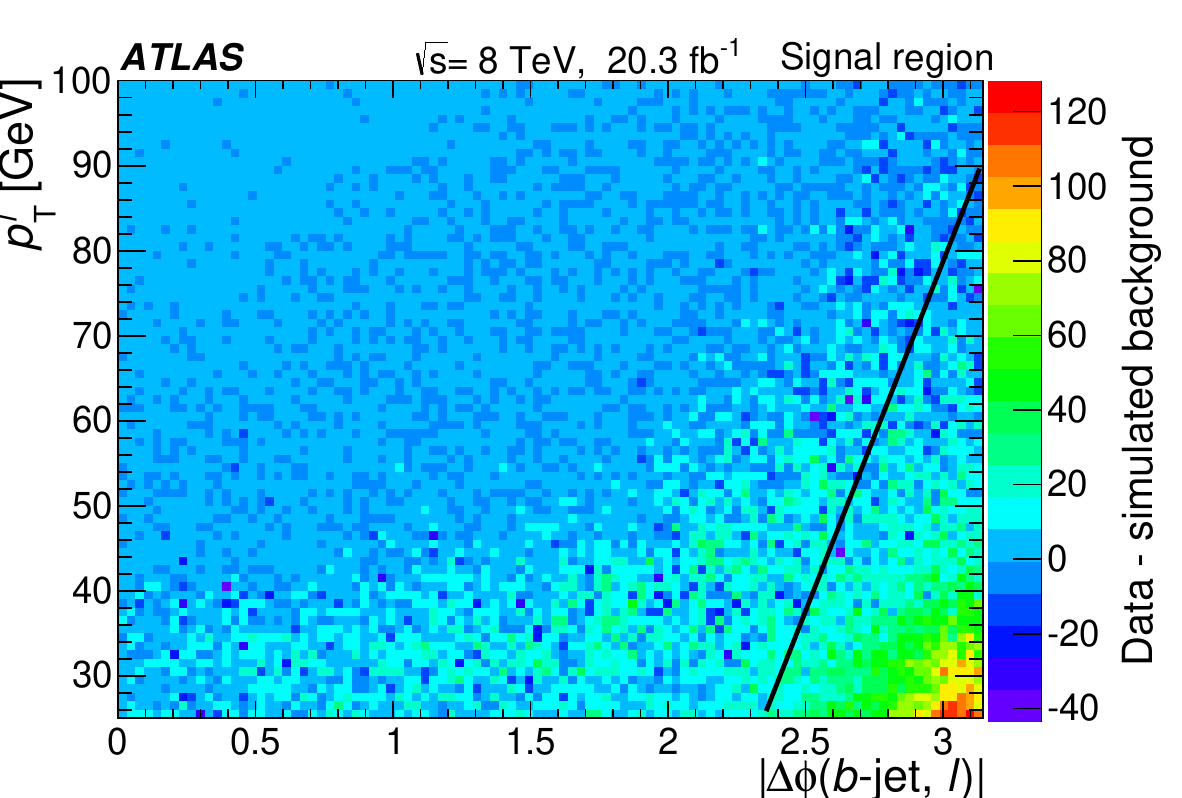}
  \caption{The transverse momentum of the lepton versus the azimuthal angle between the lepton and the jet.
    The colours indicate the number of events in data after the simulated backgrounds except the multi-jet
    contribution have been subtracted and before the cut given by \Eqn{\protect\ref{eq:antiqcd}} is applied. The solid
    black line shows the cut.
  }
  \label{fig:antiqcd}
\end{figure}

The missing transverse momentum (with magnitude \MET) is calculated 
based on the vector sum of energy deposits in the calorimeter projected
onto the transverse plane~\cite{PERF-2011-07}. All cluster energies are corrected using
the local cluster calibration scheme.
Clusters associated with a high-\pT jet or electron are further calibrated using 
their respective energy corrections. In addition, contributions from the \pT
of selected muons are included in the calculation of \MET.
Due to the presence of a neutrino in the final state of the signal process, $\MET > \SI{30}{\GeV}$ is required. 
Lepton candidates in multi-jet events typically arise from charged tracks being misidentified as leptons,
electrons arising from converted photons and leptons from $c$- and $b$-hadron decays.
Such candidates are collectively referred to as fake leptons.
As such, the multi-jet events tend to have low 
\MET and low $W$-boson transverse mass,\footnote{%
The $W$-boson transverse mass is defined as:
$\MTW = \sqrt{ 2 \left(\pT(\ell) \MET  - \vec{p}_{\text{T}}(\ell) \cdot
\vec{E}_{\text{T}}^{\text{miss}}\right)}$, where $\vec{p}_{\text{T}}(\ell)$ denotes the transverse momentum of the
lepton and $p_{\text{T}}(\ell) = |\vec{p}_{\text{T}}(\ell)|$.} \MTW, relative to single
top-quark events.
Therefore, an additional requirement on \MTW is an effective way to reduce
this background.
The selection applied is  $\MTW > \SI{50}{\GeV}$.
In order to further suppress the multi-jet background and also to remove poorly reconstructed leptons
with low transverse momentum, a requirement on the transverse momentum of leptons and the azimuthal angle between the
lepton and jet is applied:
\begin{linenomath}
\begin{equation}  \label{eq:antiqcd}
 \pT^{\ell} > \SI{90}{\GeV} \left( 1- \frac{\pi - |\Delta \phi (\ell, \text{jet})|}{\pi -2}\right)\,.
\end{equation}
\end{linenomath}
The parameters of the cut are motivated by the distribution of multi-jet events, obtained in the signal region,
where the simulated backgrounds except the multi-jet contribution are subtracted from data. 
Almost no signal events are removed by this cut. 
The distribution of the transverse momentum of the lepton versus the azimuthal angle between the lepton and the jet is shown in \Fig{\ref{fig:antiqcd}}.

In addition to the signal region defined by this selection, a control region is defined with the same kinematic requirements,
but with a less stringent $b$-tagging requirement with an efficiency of \SI{85}{\%},
and excluding events passing the tighter signal-region $b$-tagging selection.
This control region is designed such that the resulting sample is dominated by $W$+jets production, which is the dominant background. 

\subsection{Background estimation}
\label{sec:bgestimation}

For all background processes except the multi-jet background, 
the normalisations are estimated by using Monte Carlo simulation scaled to the theoretical cross-section predictions,
using $\mtop = \SI{172.5}{\GeV}$.
In order to check the modelling of kinematic distributions, correction factors to the normalisation of the $W$+jets
and \ttbar and single-top processes are subsequently determined simultaneously in the context of the multi-jet background estimation.

The SM single top-quark production cross-sections are calculated to approximate next-to-next-to-leading-order (NNLO)
precision.
The production via the $t$-channel exchange of a virtual $W$~boson has a predicted cross-section of 
\SI{87}{\pb}~\cite{Kidonakis:2011wy}. The cross-section for the associated production of an on-shell $W$ boson and 
a top quark ($Wt$ channel) has a predicted value of \SI{22.3}{\pb}~\cite{Kidonakis:2010ux}, while the 
$s$-channel production has a predicted cross-section of \SI{5.6}{\pb}~\cite{Kidonakis:2010tc}. 
The resulting weighted average of the theoretical uncertainties including PDF and scale uncertainties 
of these three processes is \SI{10}{\%}.

The cross-section of the  \ttbar process is normalised to \SI{238}{\pb},
calculated at NNLO in QCD including resummation of
next-to-next-to-leading logarithmic (NNLL) soft gluon terms~\cite{Cacciari:2011hy,Baernreuther:2012ws,Czakon:2012zr,Czakon:2012pz,Czakon:2013goa}
with Top++2.0~\cite{Czakon:2011xx}. 
The PDF and $\alpha_{\mathrm{s}}$ uncertainties are calculated using the PDF4LHC prescription~\cite{Botje:2011sn} 
with the \mstw NNLO~\cite{Martin:2009iq,Martin:2009bu} at \SI{68}{\%} \CL, 
the \ct10 NNLO~\cite{Lai:2010vv,Gao:2013xoa}, and the
\nnpdf~\cite{Ball:2012cx} PDF sets, and are added in quadrature to the scale uncertainty, 
yielding a final uncertainty of \SI{6}{\%}.

The cross-sections for inclusive $W$- and $Z$-boson production are predicted with NNLO precision 
using the \textsc{FEWZ} program~\cite{Melnikov:2006kv,Gavin:2010az}, resulting in a LO-to-NNLO $K$-factor of 1.10 
and an uncertainty of \SI{4}{\%}. The uncertainty includes the uncertainty on the PDF and scale variations.
The scale factor is applied to the prediction based on the LO \SHERPA calculation and the  
flavour composition is also taken from the MC samples. The modelling of the transverse momentum of the $W$
boson in the $W$+jets sample is improved by reweighting the simulated samples to data in the 
$W$+jets-dominated control region.

LO-to-NLO $K$-factors obtained with \MCFM~\cite{Campbell:2010ff} of the order of 1.3 are applied to the
\ALPGEN LO predictions for diboson production. 
Since the diboson process is treated together with $Z$-boson production in the statistical analysis and 
the fraction of selected events is only \SI{5}{\%}, the same uncertainties as used for the $Z$+jets process are assumed. 
 
Multi-jet events may be selected if a jet is misidentified as an isolated lepton or if the event has a non-prompt lepton 
that appears to be isolated.
The normalisation of this background is obtained from a fit to the observed \MET~distribution,
performed both in the signal and control regions.
In order to construct a sample of multi-jet background events, 
different methods are adopted for the electron and muon channels.
The \enquote{jet-lepton} model is used in the electron channel
while the \enquote{anti-muon} model is used in the muon channel~\cite{ATLAS-CONF-2014-058}.
In the jet-lepton model, a shape for the multi-jet background is established using events from a \PYTHIA
dijet sample, which are selected using same criteria as the standard selection,
but with a jet used in place of the electron candidate. Each candidate jet has to fulfil the same 
\pT and $\eta$ requirements as a standard lepton and deposit \SIrange{80}{95}{\%} of 
its energy in the electromagnetic calorimeter.
Events with an electron candidate passing the electron cuts described in \Sect{\ref{sec:objsel}} are rejected
and an event is accepted if exactly one \enquote{jet-lepton} is found.
The anti-muon model is derived from collision data.
In order to select a sample that is highly enriched with muons from multi-jet events, 
some of the muon identification cuts are inverted or changed,
e.g.\ the isolation criteria are inverted.

To determine the normalisation of the multi-jet background template, 
a binned maximum-likelihood fit is performed on the \MET~distribution using the observed data,
after applying all selection criteria except for the cut on \MET. 
Fits are performed separately in two $\eta$ regions for electrons: in the endcap ($|\eta| > 1.52$) and central ($|\eta| < 1.37$)
region of the electromagnetic calorimeter, i.e.\ the transition region is excluded.
For muons, the complete $\eta$ region is used.
The multi-jet templates for both the electrons and the muons are fitted together with templates derived from MC
simulation for all other background processes (top  quark, $W$+light flavour (LF), $W$+heavy flavour (HF), $Z$+jets, dibosons).
Acceptance uncertainties are accounted for in the fitting process
in the form of additional constrained nuisance parameters. 
For the purpose of these fits, the contributions from \WLF and \WHF,
the contributions from $t\bar{t}$ and single top-quark production,
and the contributions from $Z$+jets and diboson production are each combined into one template.
The normalisation of the template for $Z$+jets and diboson production is fixed during the fit, 
as its contribution is very small.

\begin{figure}[htbp]
  \centering
  \subfigure[][]{\label{subfig:missetfit_CR_ele}%
     \includegraphics[width=0.45\textwidth]{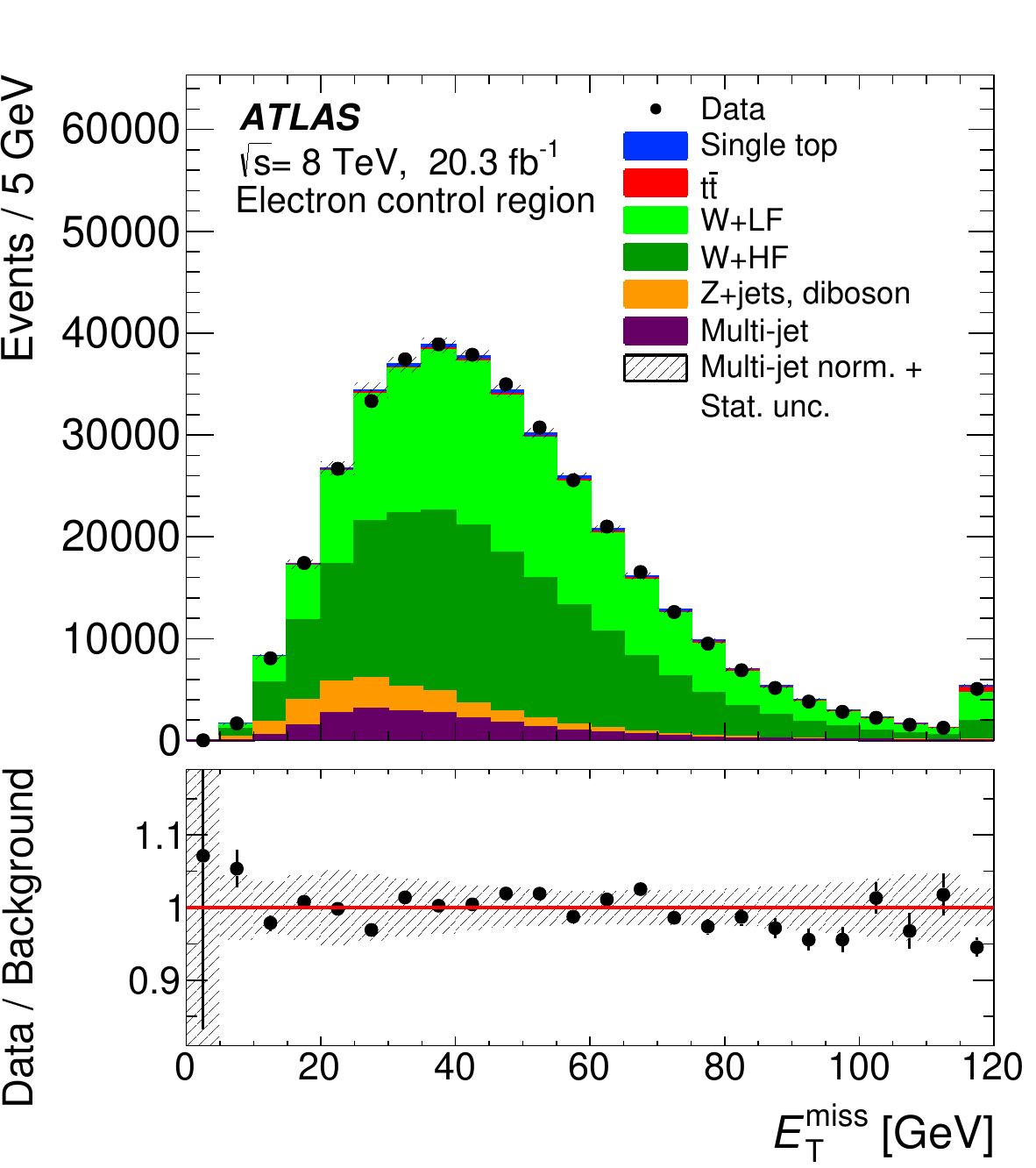}
  }
  \quad
  \subfigure[][]{\label{subfig:missetfit_CR_muo}%
     \includegraphics[width=0.45\textwidth]{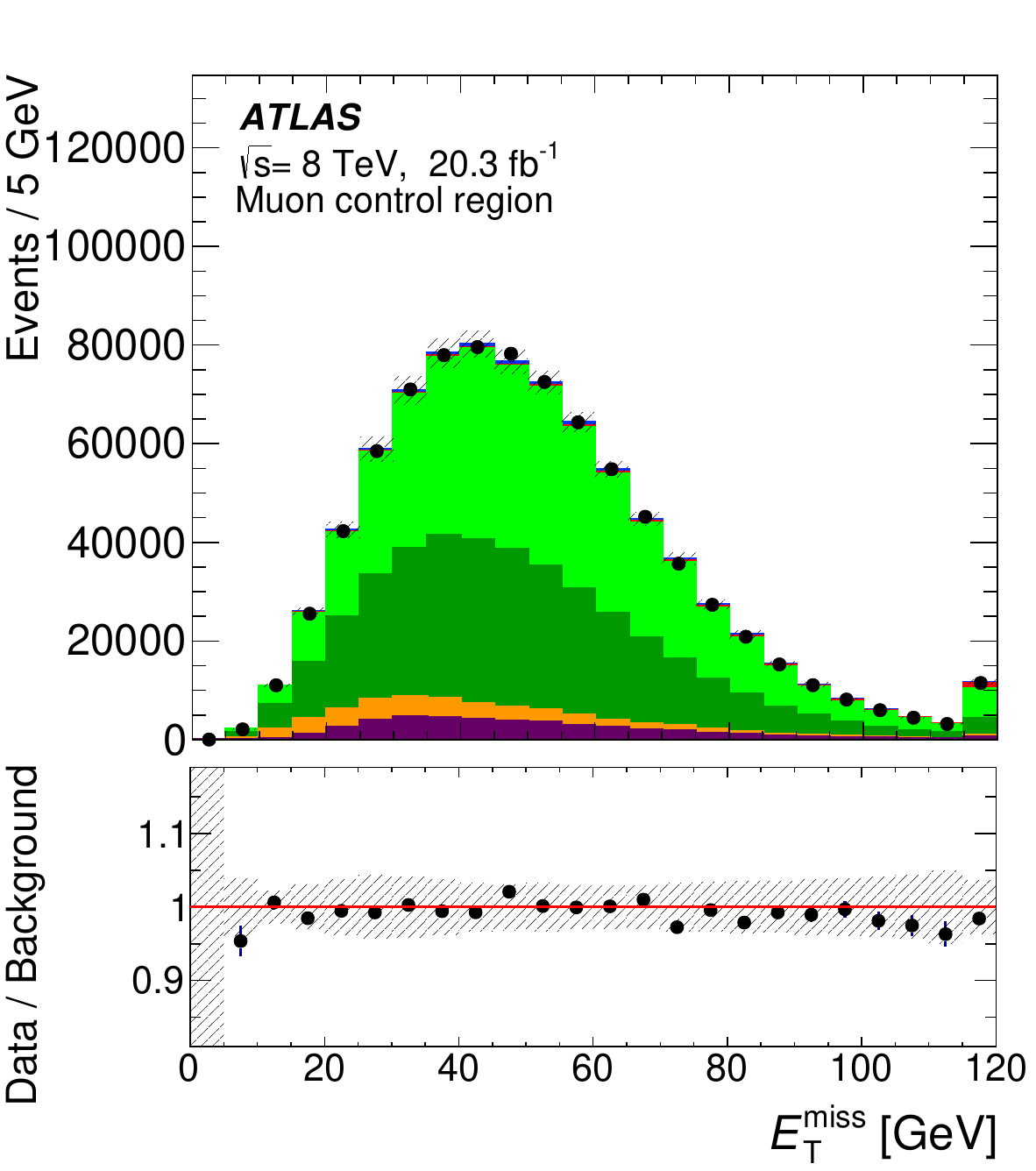}
  }\\
   
  \subfigure[][]{\label{subfig:missetfit_SR_ele}%
     \includegraphics[width=0.45\textwidth]{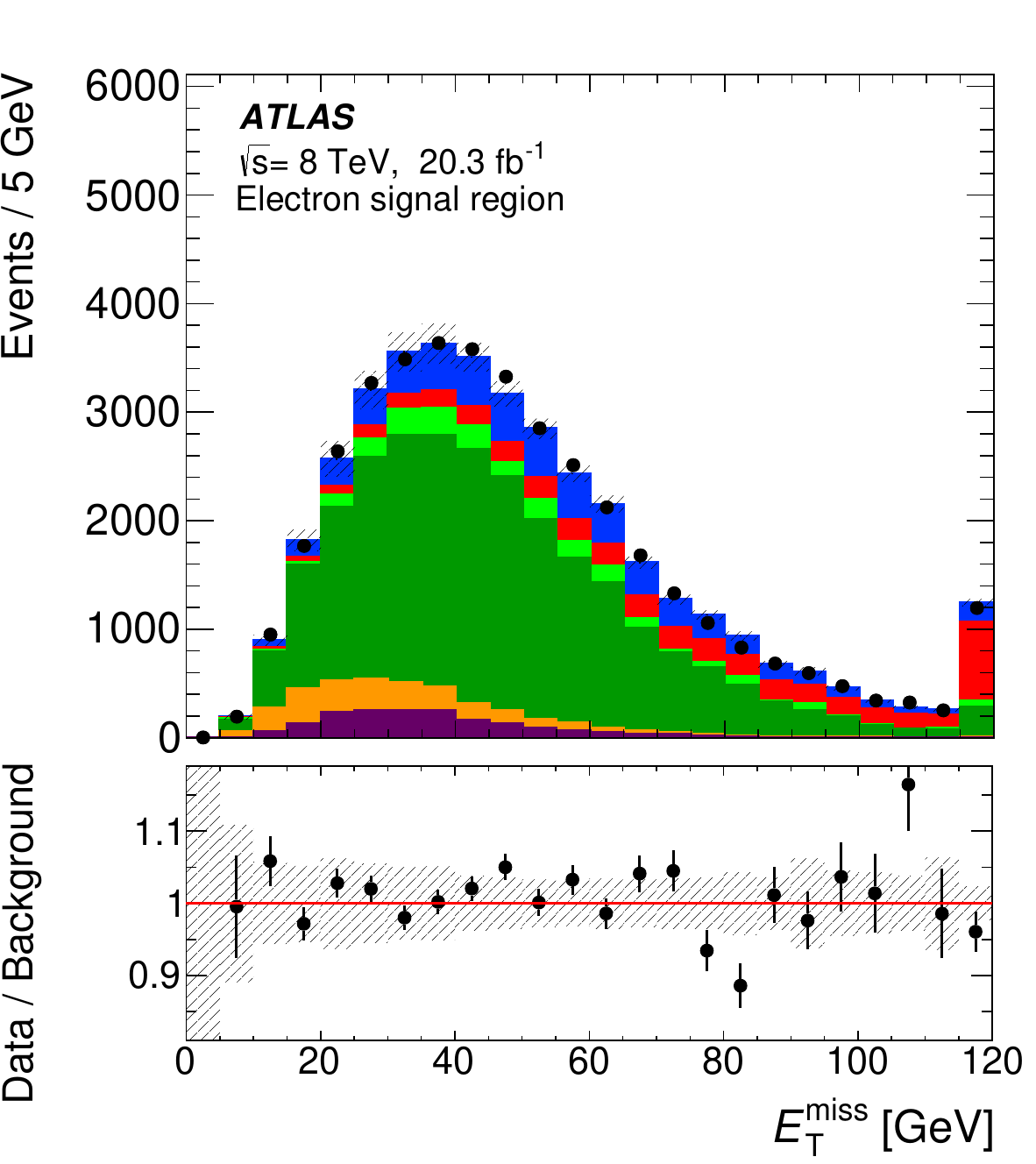}
  }
  \quad
  \subfigure[][]{\label{subfig:missetfit_SR_muo}%
     \includegraphics[width=0.45\textwidth]{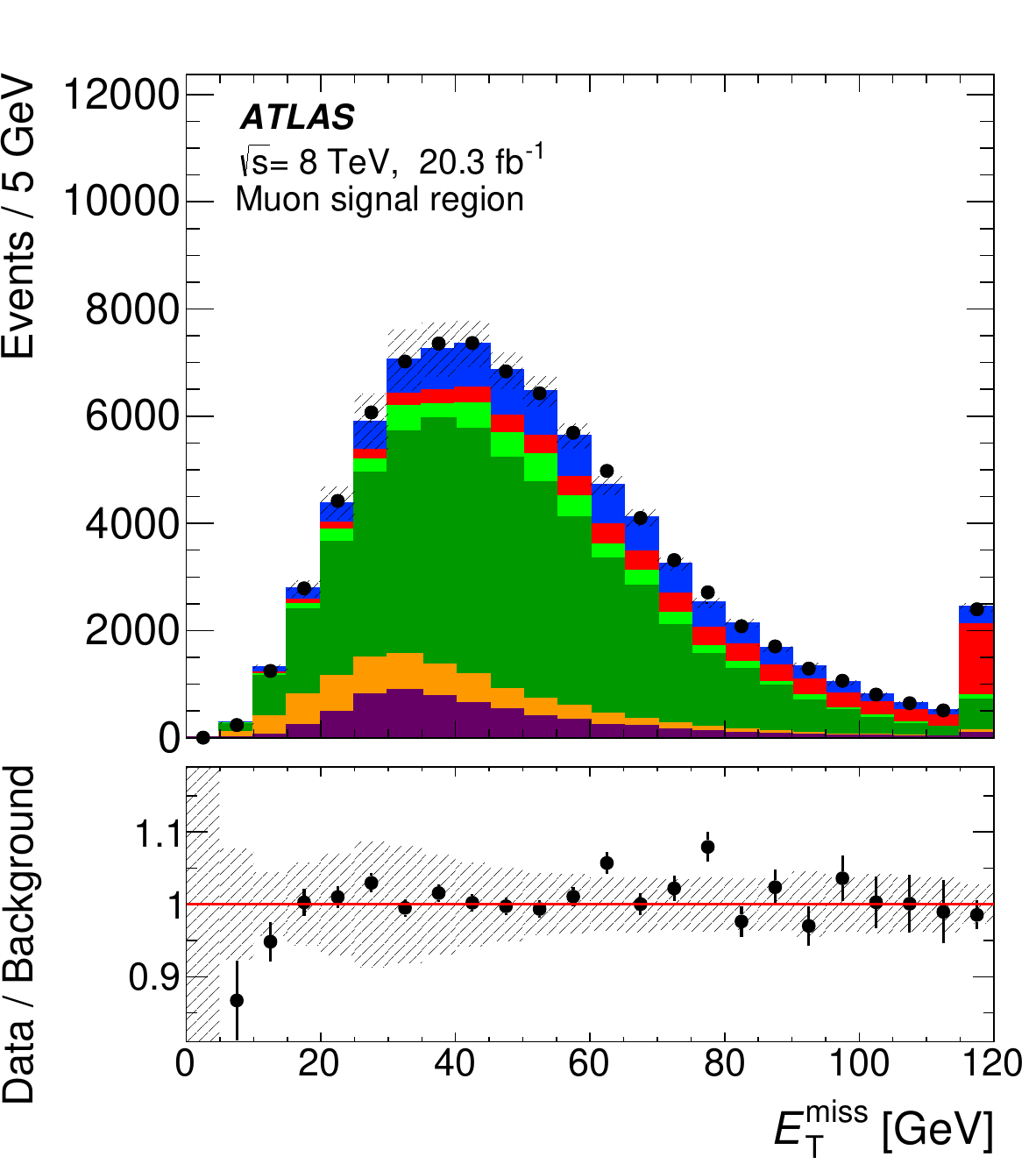}
  }
  
  \caption{Fitted distributions of the missing transverse momentum $\MET$ for \subref{subfig:missetfit_CR_ele}~central
  electrons and \subref{subfig:missetfit_CR_muo}~muons in the control region and for \subref{subfig:missetfit_SR_ele}~central electrons
    and \subref{subfig:missetfit_SR_muo}~muons in the signal region. 
    The last histogram bin includes overflow events and the hatched error bands contain 
    the MC statistical uncertainty combined with the normalisation uncertainty on the multi-jet
    background.
    }
  \label{fig:missetfit}
\end{figure}

The \MET~distributions after rescaling the different backgrounds and the multi-jets template to their 
respective fit results are shown in \Fig{\ref{fig:missetfit}} for both the electron and the muon channels.
The fitted scale factors for the other templates are close to 1.

Table~\ref{tab:evtyield} provides the event yields after the complete event selection
for the control and signal regions.
The yields are calculated using the
acceptance from MC samples normalised to their respective theoretical cross-sections including the (N)NLO $K$-factors,
while the number of expected events for the multi-jet background is obtained from the maximum-likelihood fit.
Each event yield uncertainty combines the statistical uncertainty, 
originating from the limited size of the simulation samples, 
with the uncertainty on the cross-section or normalisation.
The observed event yield in data agrees well with the background prediction.
For comparison, a \SI{1}{\pb} FCNC cross-section would lead to 530 events in the signal region.
The corresponding efficiency for selecting FCNC events is \SI{3.1}{\%}.

\begin{table}[htbp]
  \sisetup{round-mode=figures}
  \centering
   \begin{tabular}{l S[table-format=7.0,round-precision=3]@{$\,\pm\,$}S[table-format=6.0]
                     S[table-format=6.0,round-precision=3]@{$\,\pm\,$}S[table-format=5.0]}
    \toprule
        Process           & \multicolumn{2}{c}{Control region} & \multicolumn{2}{c}{Signal region} \\
    \midrule
        Single top        &     11506 & 615   &  14374 & 769  \\ 
        \ttbar            &     10658 & 654   &  11962 & 735  \\ 
        $W$+LF            &   \multicolumn{1}{r@{$\,\pm\,$}}{\num[round-precision=4]{526877}} & 126633  &  \multicolumn{1}{r@{$\,\pm\,$}}{\num[round-precision=2]{6711}} & 1884 \\ 
        $W$+HF            &   \multicolumn{1}{r@{$\,\pm\,$}}{\num[round-precision=4]{445206}} & 244924  & 62121 & 34194 \\ 
        $Z$+jets          &    40039 & 9696  &  4991 & 1218  \\ 
        Multi-jet         &    68305 & 11611  &  7431 &  1263 \\
        \midrule
        Total expected    &   1102591 & 276140 & 107490 & 34308 \\
    \bottomrule
        Data              &   \multicolumn{2}{l}{\num[round-mode=off]{1112225}} & \multicolumn{2}{l}{\num[round-mode=off]{108152}} \\
    \bottomrule
  \end{tabular}
  \caption{Number of observed and expected events in the control and signal region for all lepton categories added
    together. The uncertainties shown are derived using the statistical uncertainty from the limited size of the samples and the uncertainty
    on the theoretical cross-section only or multi-jet normalisation.
    The scale factors obtained from the multi-jet background fit are not applied when determining the expected number of events.}
  \label{tab:evtyield}
\end{table}

\begin{figure}[htbp]
  \centering
  \subfigure[][]{\label{subfig:CR_leppt}%
    \includegraphics[width=0.32\textwidth]{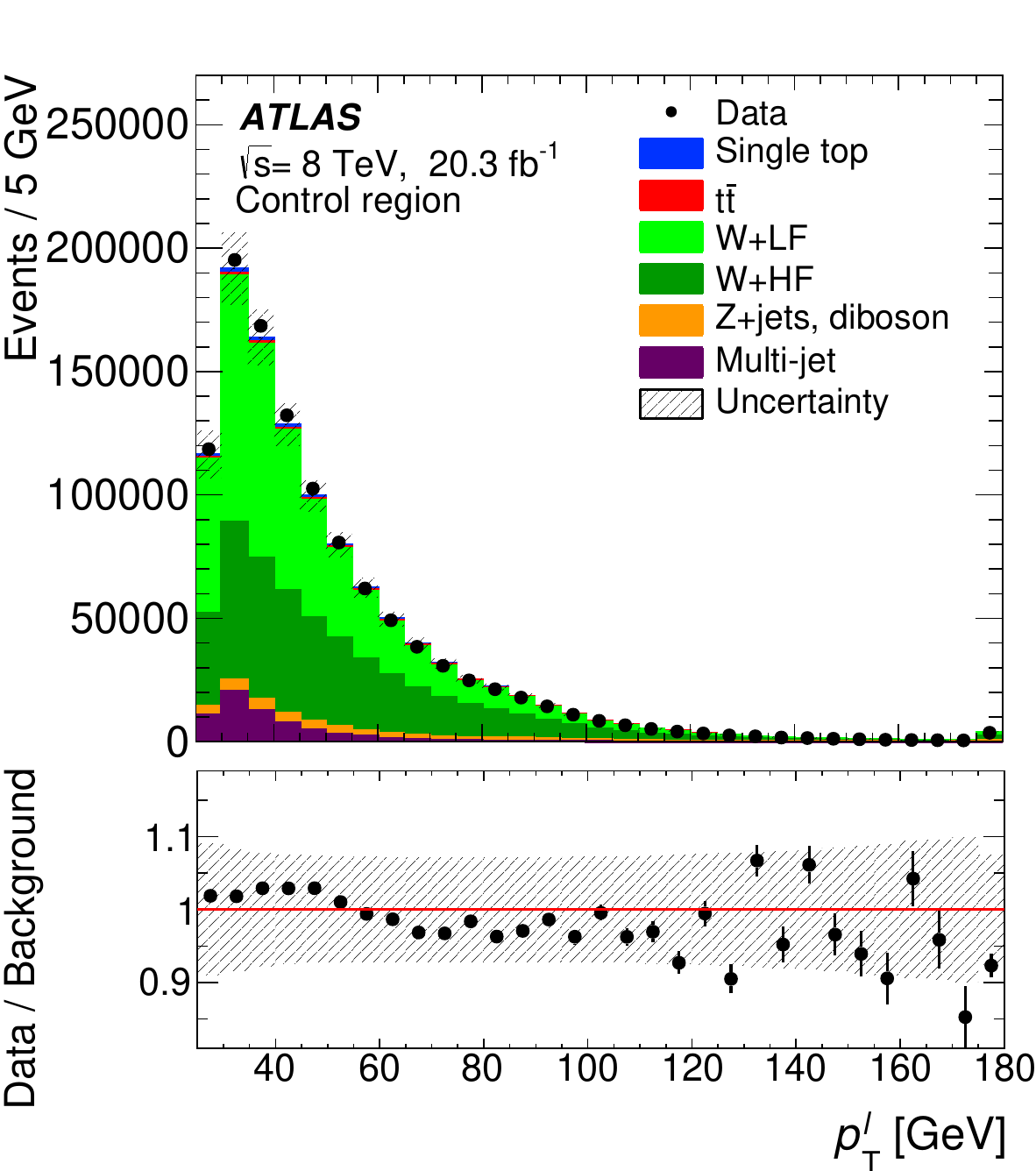}
  }
  \quad
  \subfigure[][]{\label{subfig:CR_lepeta}%
    \includegraphics[width=0.32\textwidth]{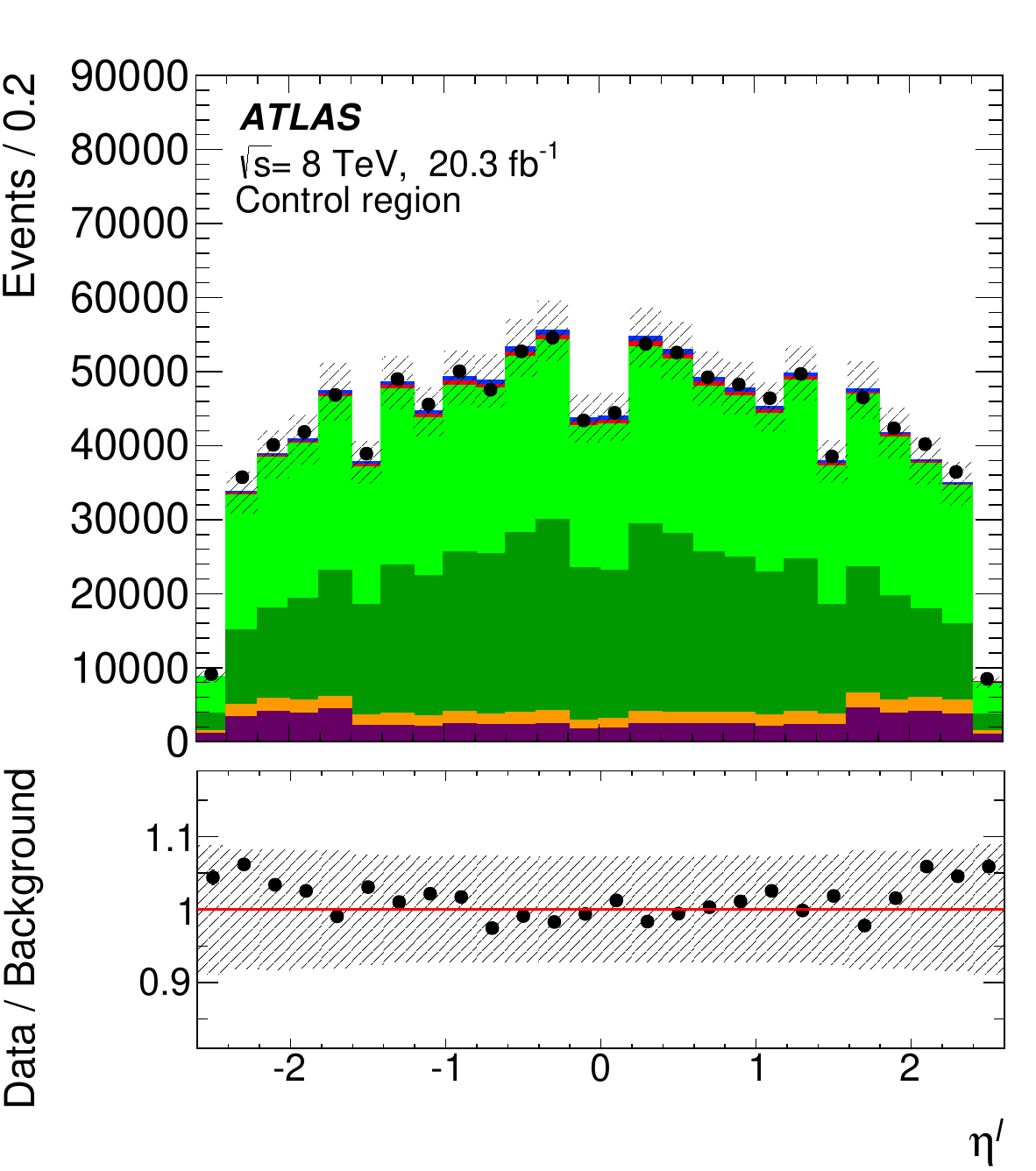}
  }\\
  \subfigure[][]{\label{subfig:CR_jetpt}%
     \includegraphics[width=0.32\textwidth]{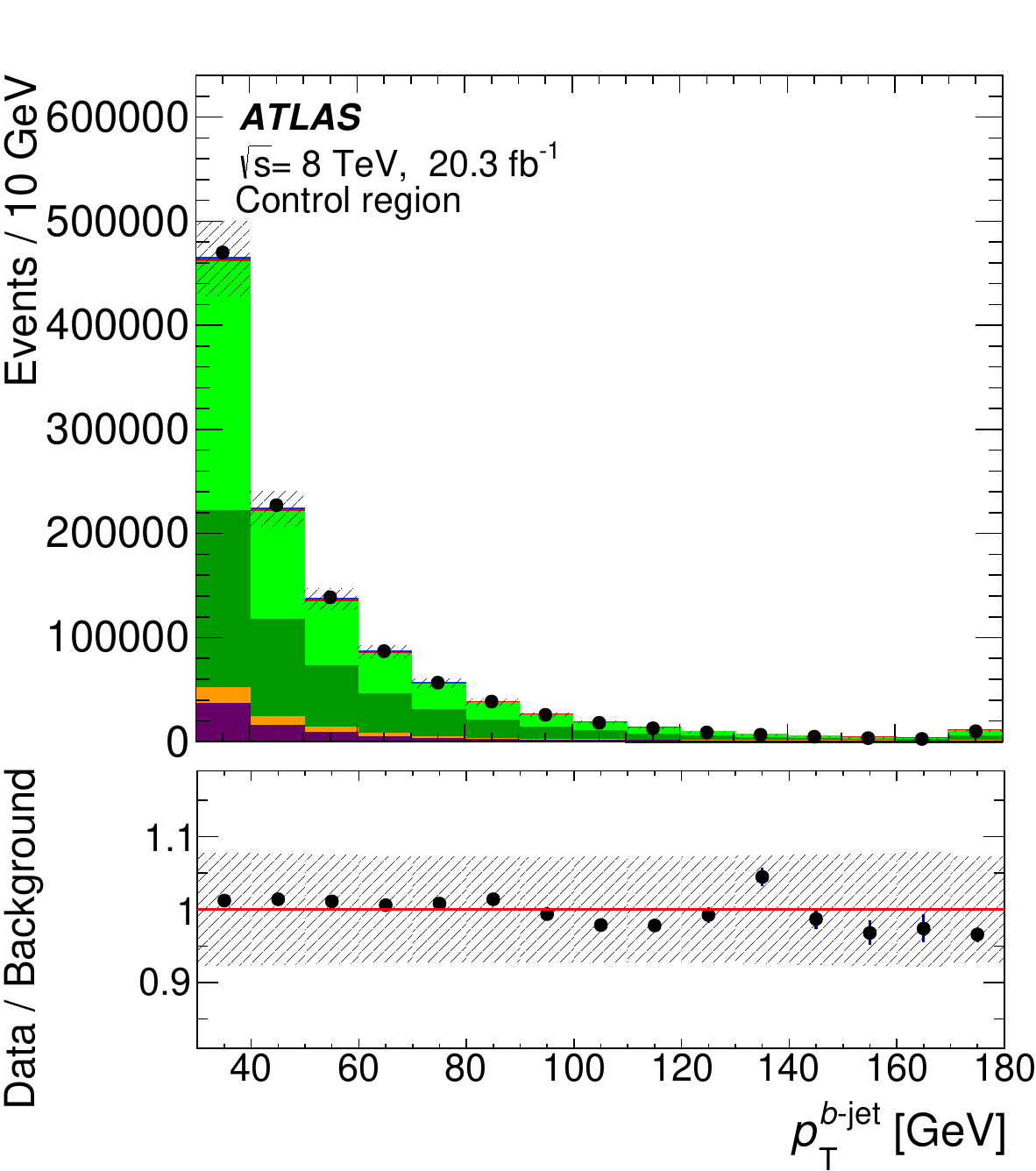}
}        
  \subfigure[][]{\label{subfig:CR_jeteta}%
     \includegraphics[width=0.32\textwidth]{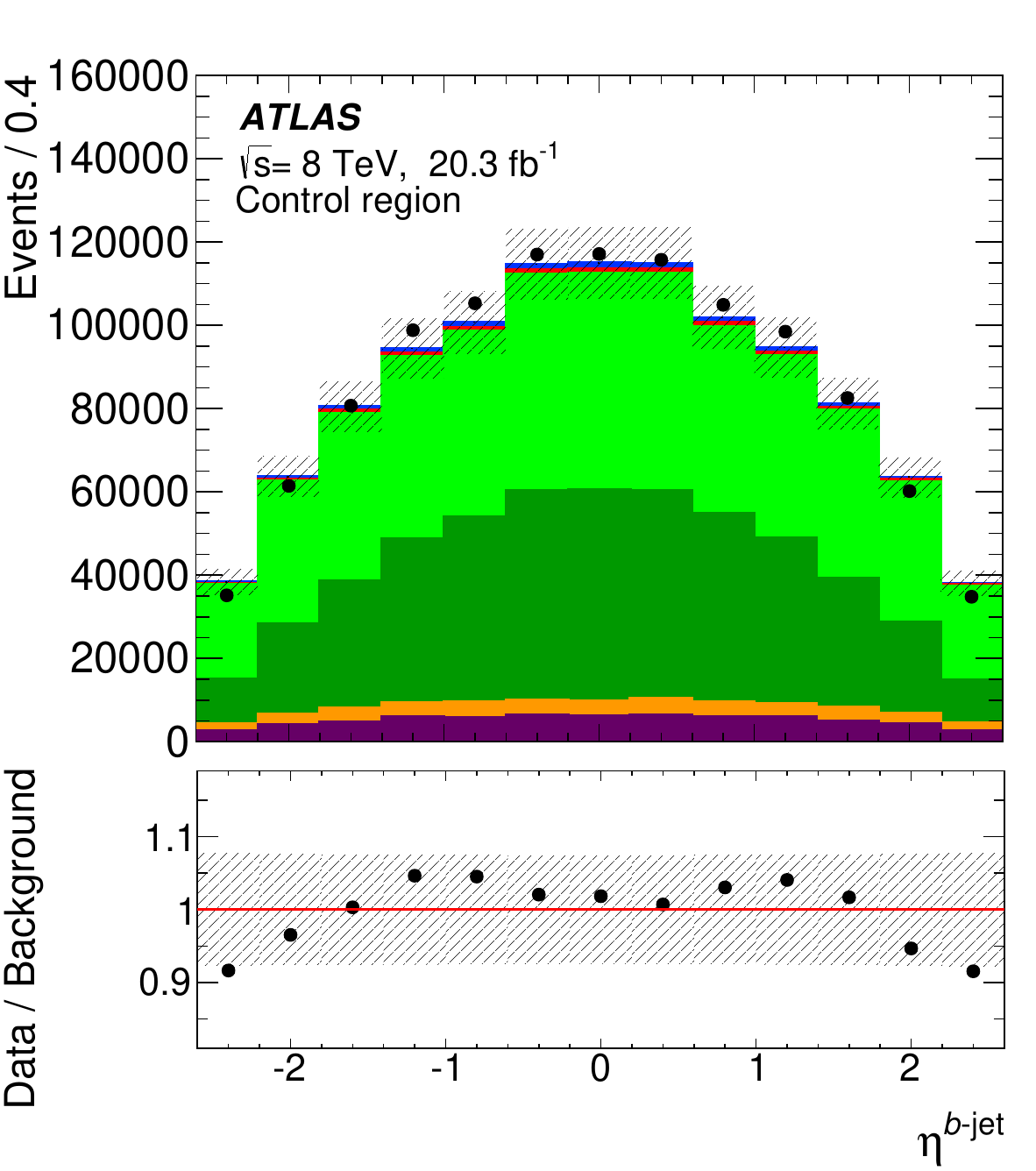}
  }\\
  \subfigure[][]{\label{subfig:CR_met}%
     \includegraphics[width=0.32\textwidth]{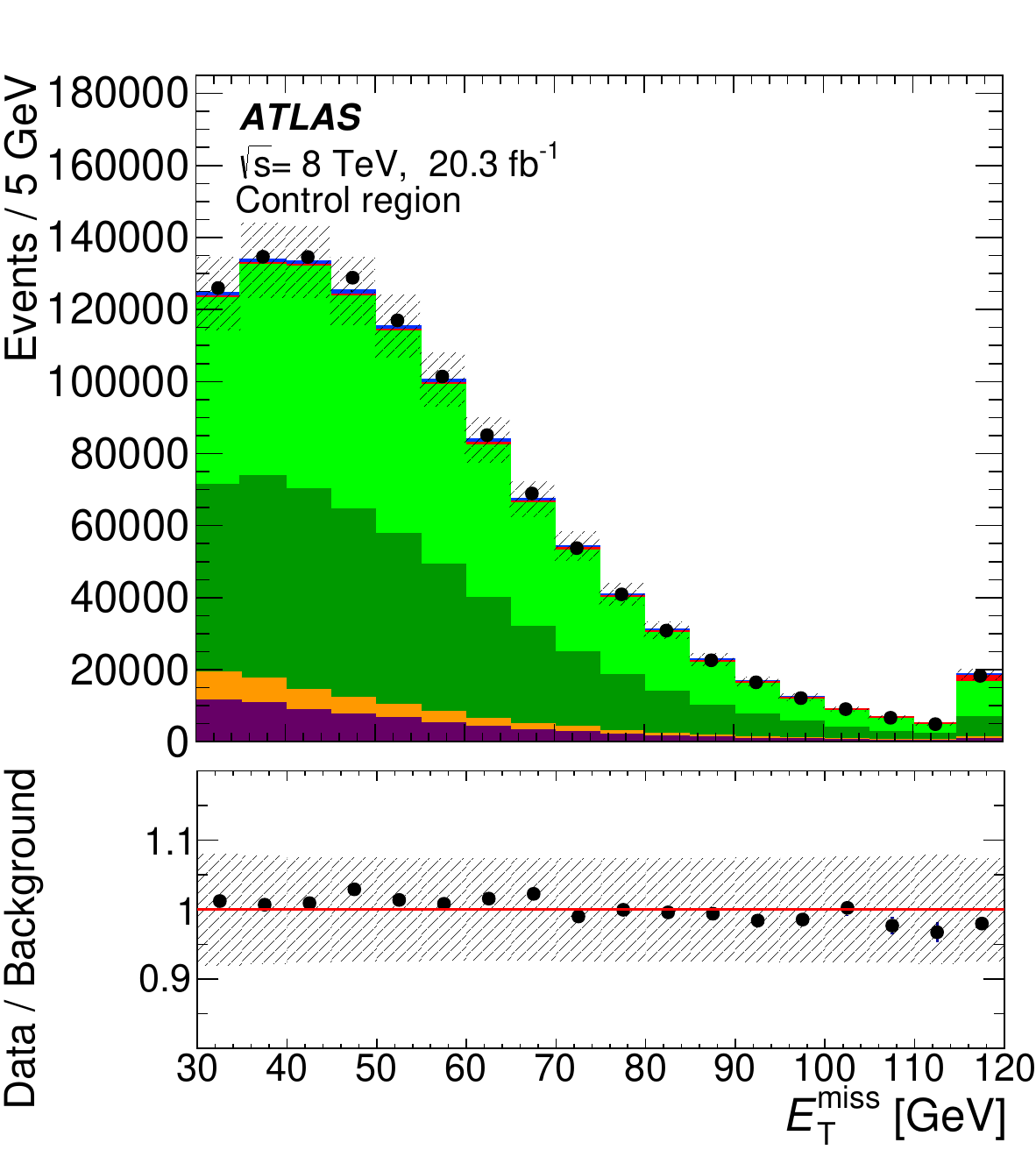}
  }
  \quad
  \subfigure[][]{\label{subfig:CR_mtw}%
     \includegraphics[width=0.32\textwidth]{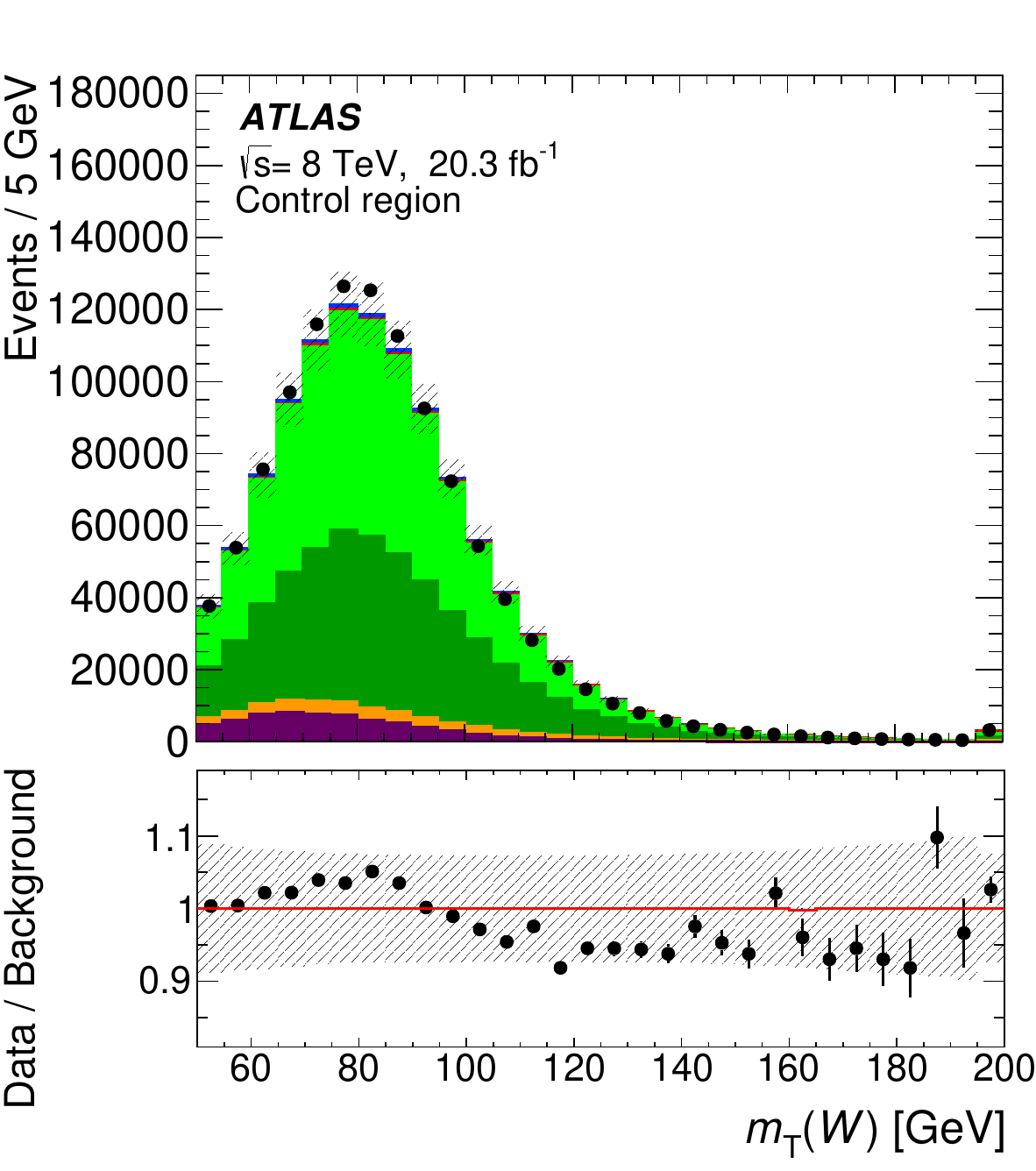}
  }

  \caption{Kinematic distributions in the control region for the combined electron and muon channels.
      All processes are normalised to the result of the binned maximum-likelihood fit used to determine the fraction of
      multi-jet events.
      Shown are: \subref{subfig:CR_leppt} the transverse momentum and \subref{subfig:CR_lepeta} pseudorapidity of
      the lepton, \subref{subfig:CR_jetpt} the transverse momentum and \subref{subfig:CR_jeteta} pseudorapidity
      of the jet, \subref{subfig:CR_met} the missing transverse momentum and \subref{subfig:CR_mtw} $W$-boson
      transverse mass.
      The last histogram bin includes overflow events and the hatched band indicates the combined statistical and systematic uncertainties, evaluated after the fit discussed in Section \ref{sec:result}.
      }
  \label{fig:control}
\end{figure}

Kinematic distributions in the control region of the identified lepton, reconstructed jet, \MET and \MTW 
are shown in \Fig{\ref{fig:control}} for the combined electron and muon channels.
These distributions are normalised using the scale factors obtained in the \MET fit to estimate the multi-jet background.
Overall, good agreement between the observed and expected distributions is seen.
The trends that can be seen in some of the distributions are covered by the systematic uncertainties.

\section{Analysis strategy}
\label{sec:analysis}

As no single variable provides sufficient discrimination between signal and background events
and the separation power is distributed over many correlated variables, 
multivariate analysis techniques are necessary to separate signal candidates from background candidates.
A neural-network (NN) classifier~\cite{Feindt:2006pm} that combines a three-layer feed-forward neural network 
with a preprocessing of the input variables is used.
The network infrastructure consists of one input node for each input variable plus one bias node,
an arbitrary number of hidden nodes, and one output node, which gives a continuous output in the interval $[-1,1]$.
The training is performed with a mixture of \SI{50}{\%} signal and \SI{50}{\%} background events, 
where the different background processes are weighted according to their number of expected events.
Only processes from simulated events are considered in the training, i.e.\ no multi-jet events are used.
In order to check that the neural network is not overtrained, \SI{20}{\%} of the available simulated events are used as
a test sample.
Subsequently, the NN classifier is applied to all samples.

The $qg\to t\to b\ell\nu$ process is characterised by three main differences from SM processes.
Firstly, the \pT distribution of the top quark is much softer 
than the \pT distribution of top quarks produced through SM top-quark production,
since the top quark is produced almost without transverse momentum. 
Hence, the $W$ boson and $b$-quark from the top-quark decay are produced almost back-to-back in the transverse plane. 
Secondly, unlike in the $W/Z$+jets and diboson backgrounds, the $W$ boson from the top-quark decay 
has a high momentum and its decay products tend to have small angles.
Lastly, the top-quark charge asymmetry differs between FCNC processes and SM processes in the $ugt$ channel. 
In $pp$ collisions, the FCNC processes are predicted to produce four times more single top quarks than anti-top quarks,
whereas in SM single top-quark production and in all other SM backgrounds this ratio is at most two.
Several categories of variables are considered as potential discriminators between the signal and 
background processes. Apart from basic event kinematics such as the \MTW or $H_{\text{T}}$ 
(the scalar sum of the transverse momenta of all objects in the final state), various object combinations 
are considered as well. 
These include the basic kinematic properties of reconstructed objects like the $W$ boson and the top quark, 
as well as angular distances in $\eta$ and $\phi$ between the reconstructed and final-state objects in the 
laboratory frame and in the rest frames of the $W$ boson and the top quark.
In order to reconstruct the four-vector of the $W$ boson, a mass constraint is used.
A detailed description of the top-quark reconstruction is given in Ref.~\cite{TOPQ-2012-21}.
Further, integer variables such as the charge of the lepton are considered.

\begin{figure}[htbp]
  \centering
  \subfigure[][]{\label{subfig:Shape_mttop}%
    \includegraphics[width=0.32\textwidth, trim = 0 75 0 0]{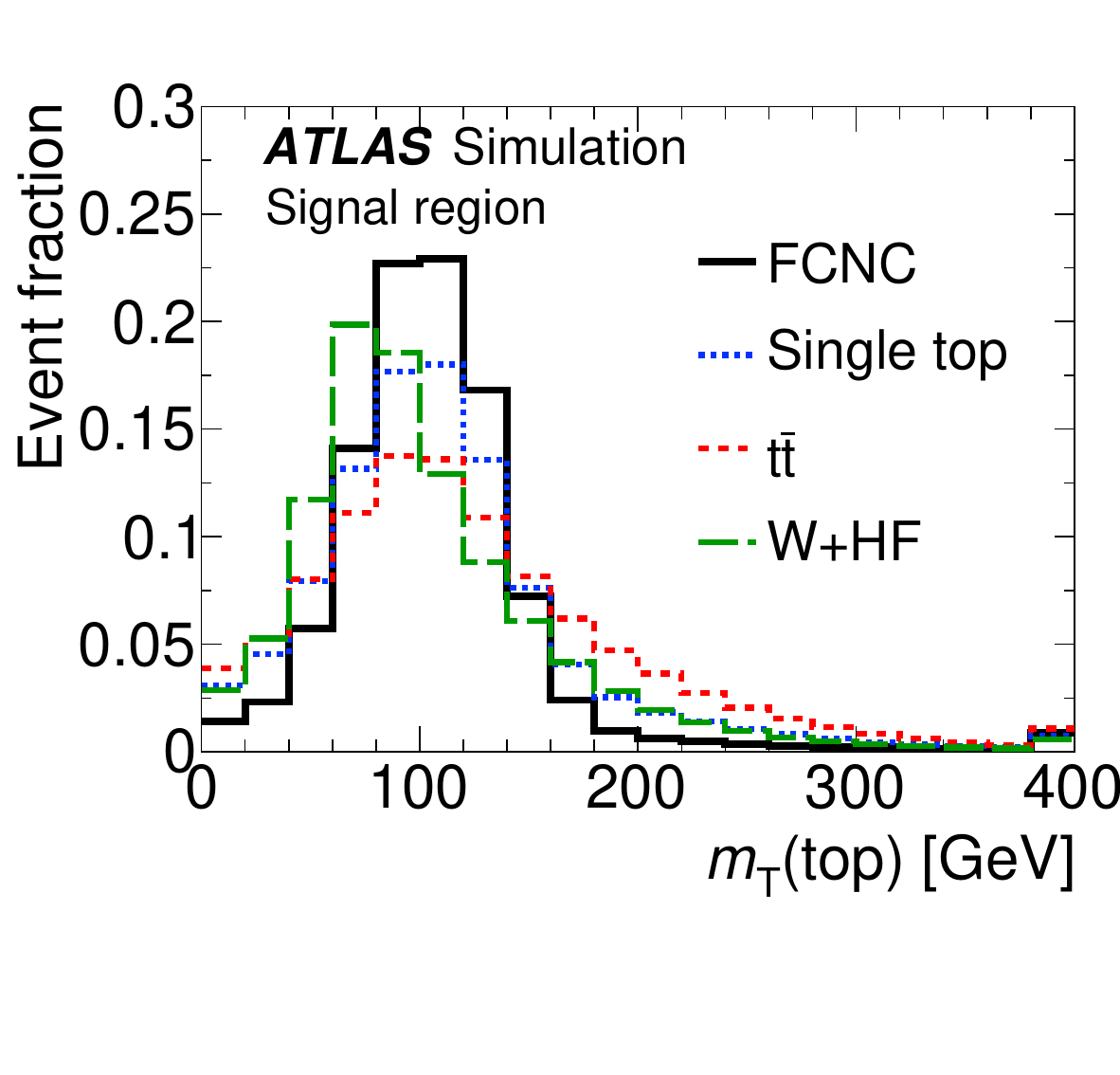}

  }
  \subfigure[][]{\label{subfig:Shape_ptlep}%
    \includegraphics[width=0.32\textwidth, trim = 0 75 0 0]{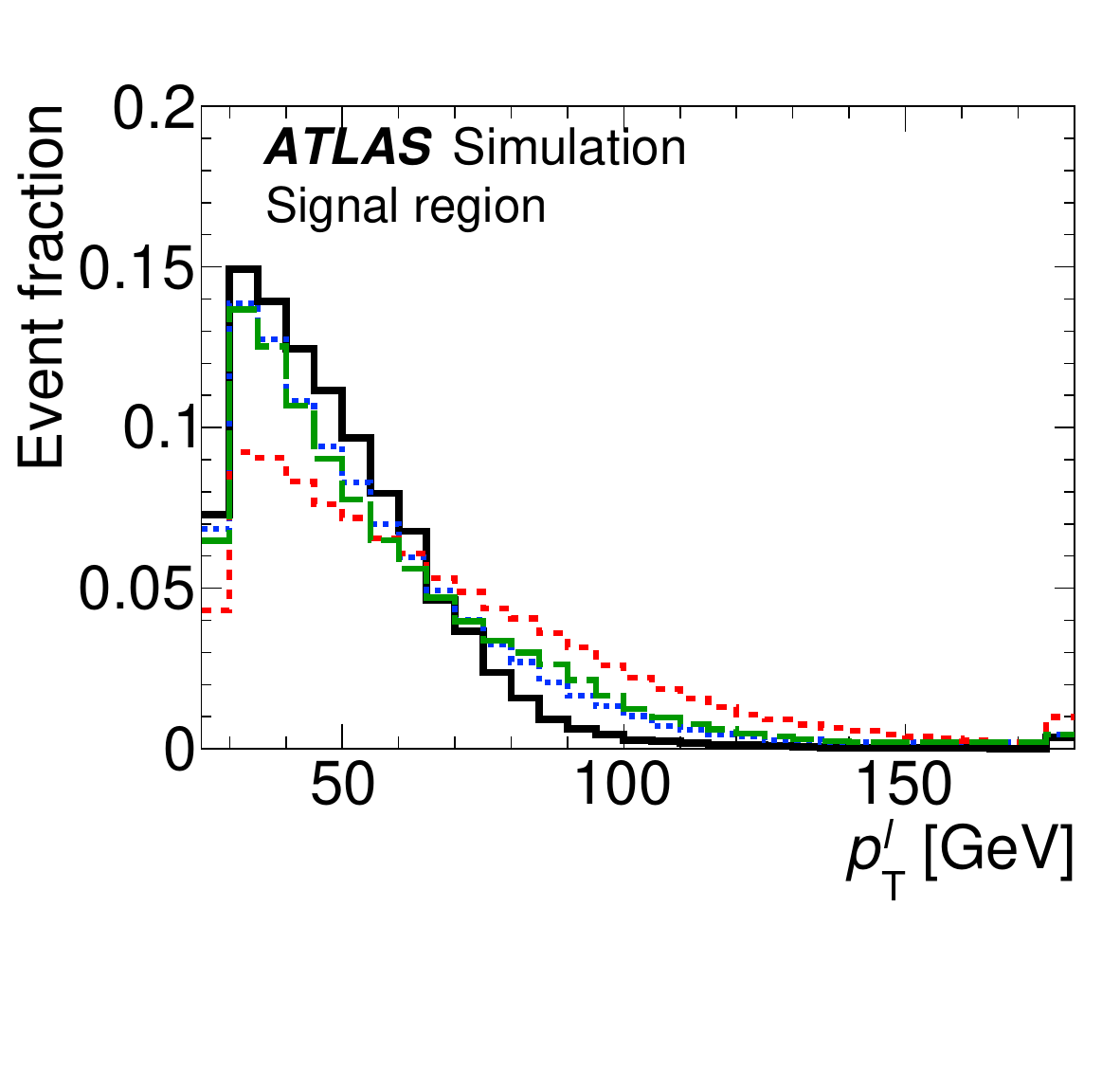}
  }
  \subfigure[][]{\label{subfig:Shape_drleptop}%
     \includegraphics[width=0.32\textwidth, trim = 0 75 0 0]{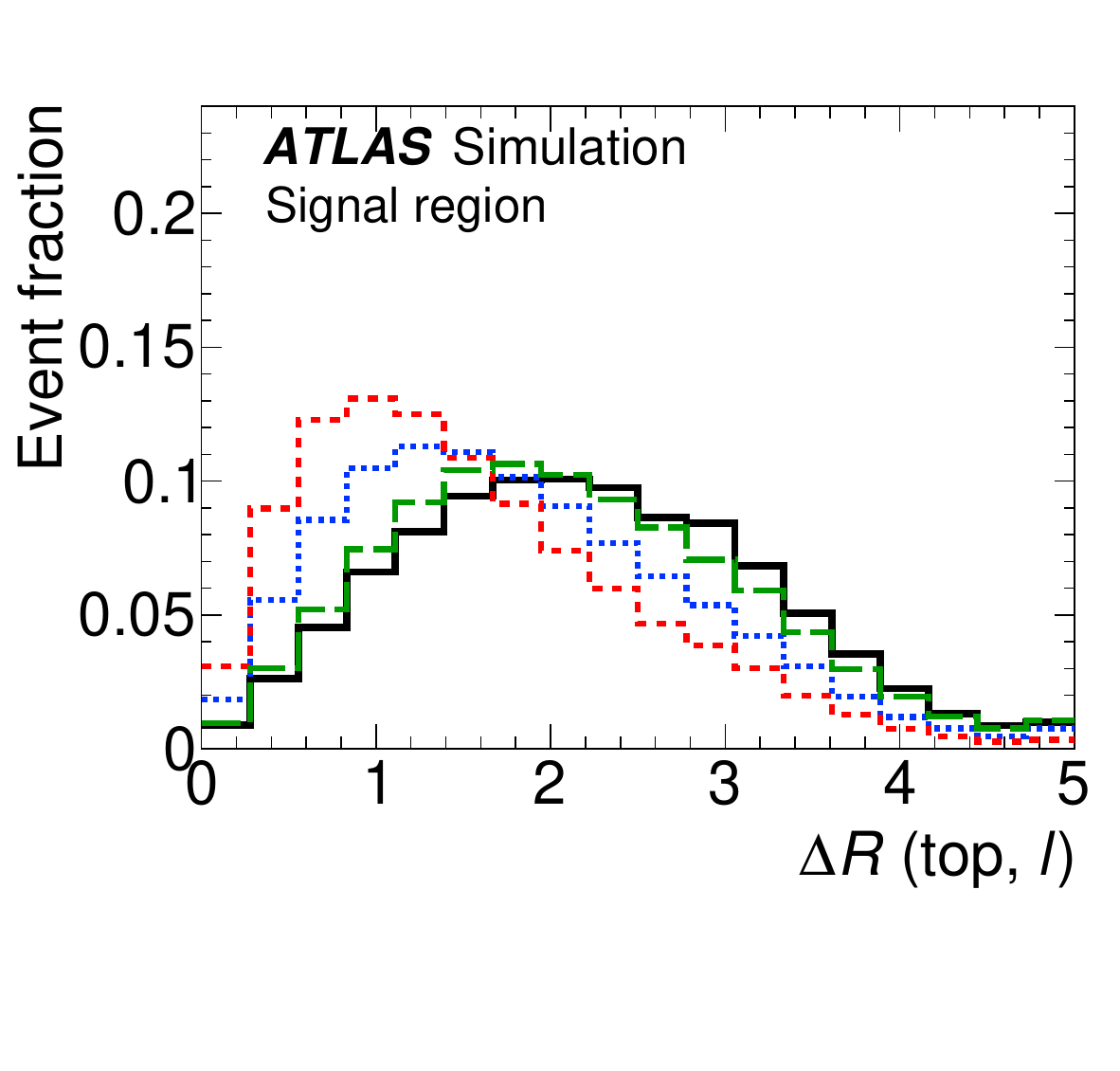}
  }
  \caption{Probability densities of the three most important discriminating variables:
    \subref{subfig:Shape_mttop} the transverse mass of the reconstructed top quark; 
    \subref{subfig:Shape_ptlep} the transverse momentum of the charged lepton; and 
    \subref{subfig:Shape_drleptop} the distance in the $\eta$--$\phi$ plane between the charged lepton and the
    reconstructed top quark.
    The last histogram bin includes overflows.}
  \label{fig:bestthreenorm}
\end{figure}

The ranking of the variables in terms of their discrimination power is automatically determined as part of the
preprocessing step and is independent of the training procedure~\cite{TOPQ-2011-14}.\footnote{%
The ranking is done according to the correlation to the output.}
Only the highest-ranking variables are chosen for the training of the neural network.
Each variable is tested beforehand for agreement between the background model and the distribution of the observed events in 
the control region.
Using only variables with an \textit{a priori} defined separation power, 13 variables remain in the network.
\Tab{\ref{tab:trainingVars}} shows a summary of the variables used, ordered by their importance.
The probability density of the three most important discriminating variables 
for the dominant background processes together with the signal is displayed in \Fig{\ref{fig:bestthreenorm}}.

\begin{figure}[htbp]
  \centering
  \subfigure[][]{\label{subfig:CR_mttop}%
    \includegraphics[width=0.32\textwidth]{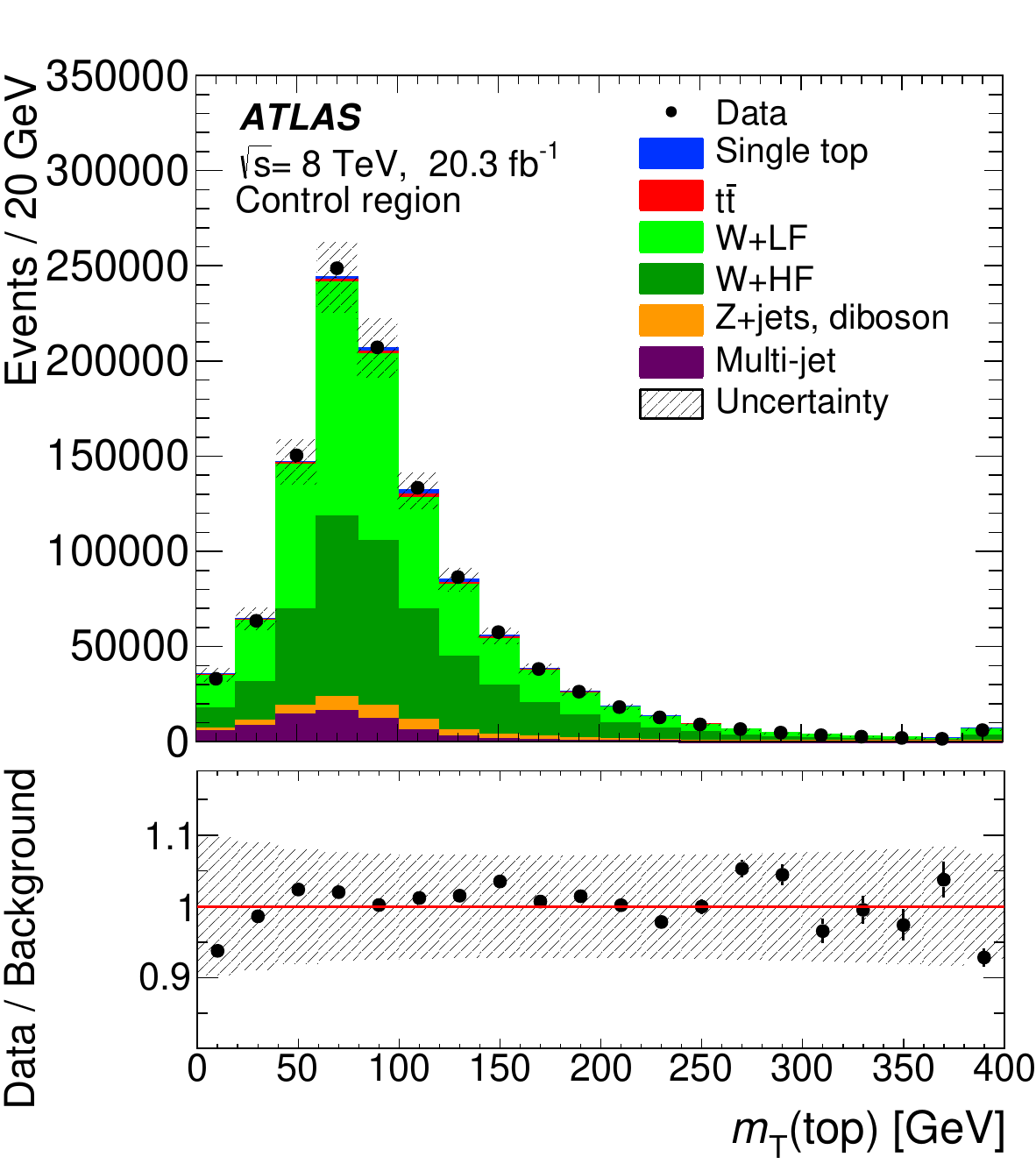}
  }
  \subfigure[][]{\label{subfig:CR_drleptop}%
    \includegraphics[width=0.32\textwidth]{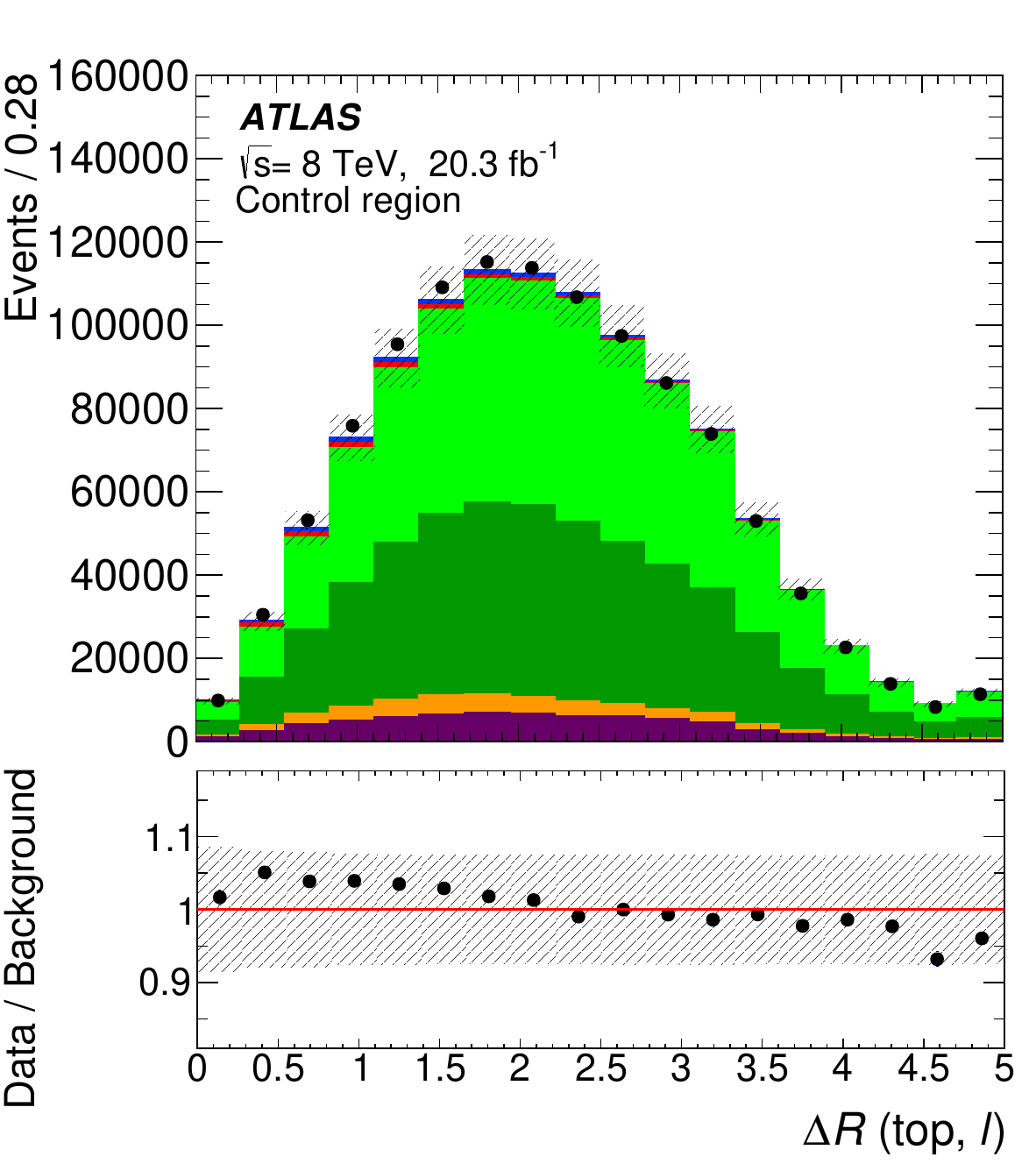}
  }
  \subfigure[][]{\label{subfig:CR_dphijettop}%
    \includegraphics[width=0.32\textwidth]{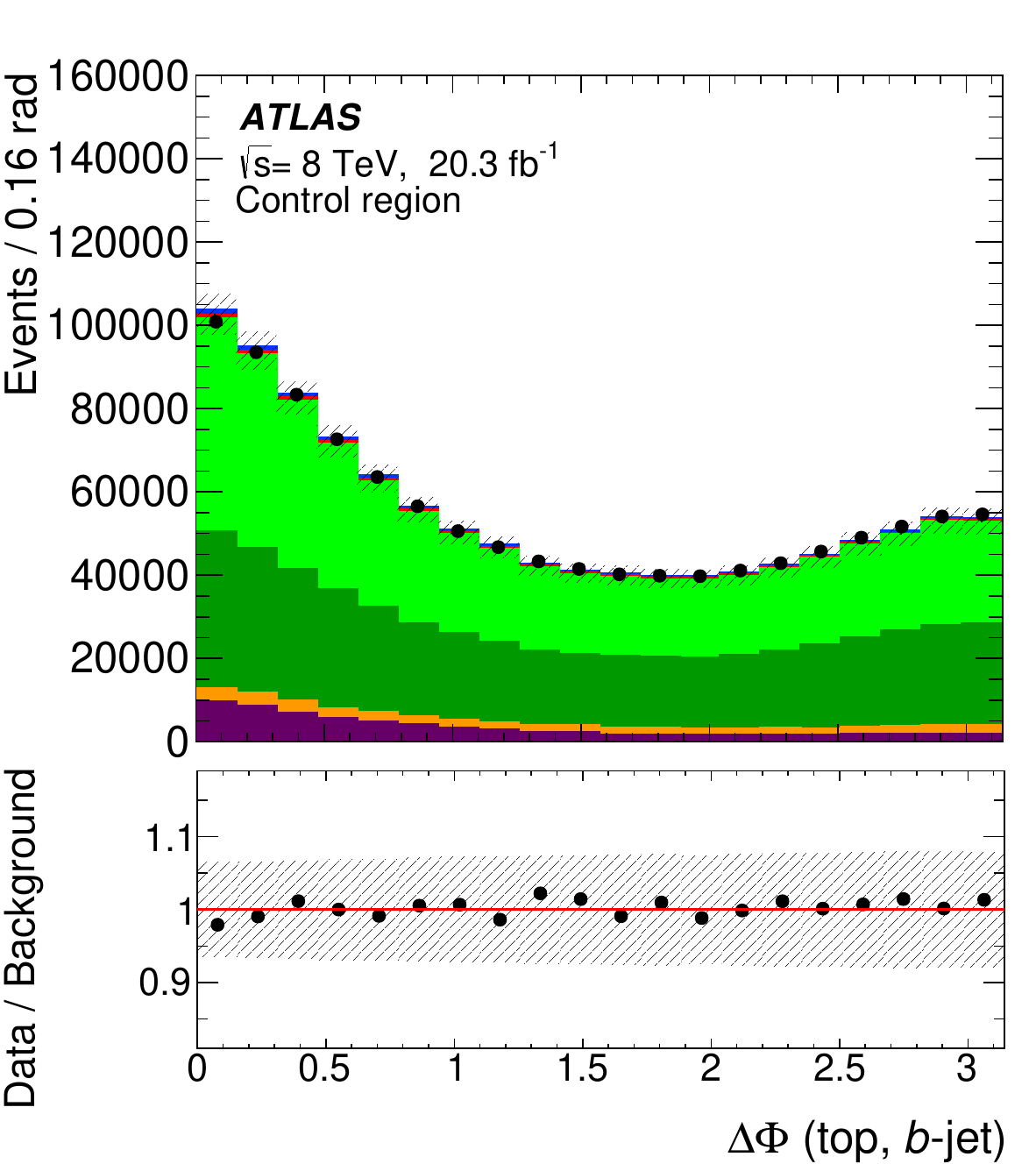}
  }\\

  \subfigure[][]{\label{subfig:SR_mttop}%
    \includegraphics[width=0.32\textwidth]{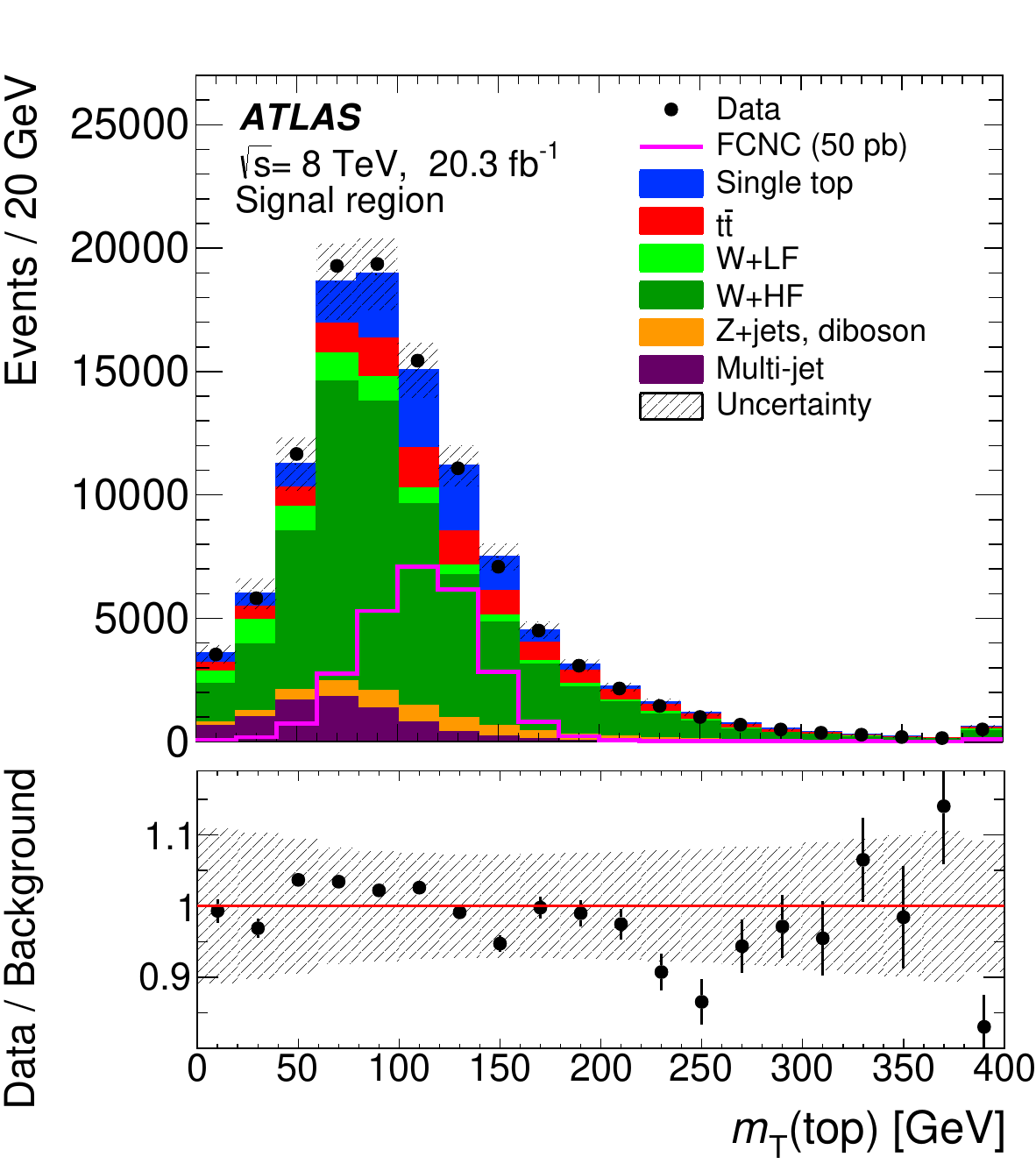}
  }
  \subfigure[][]{\label{subfig:SR_drleptop}%
    \includegraphics[width=0.32\textwidth]{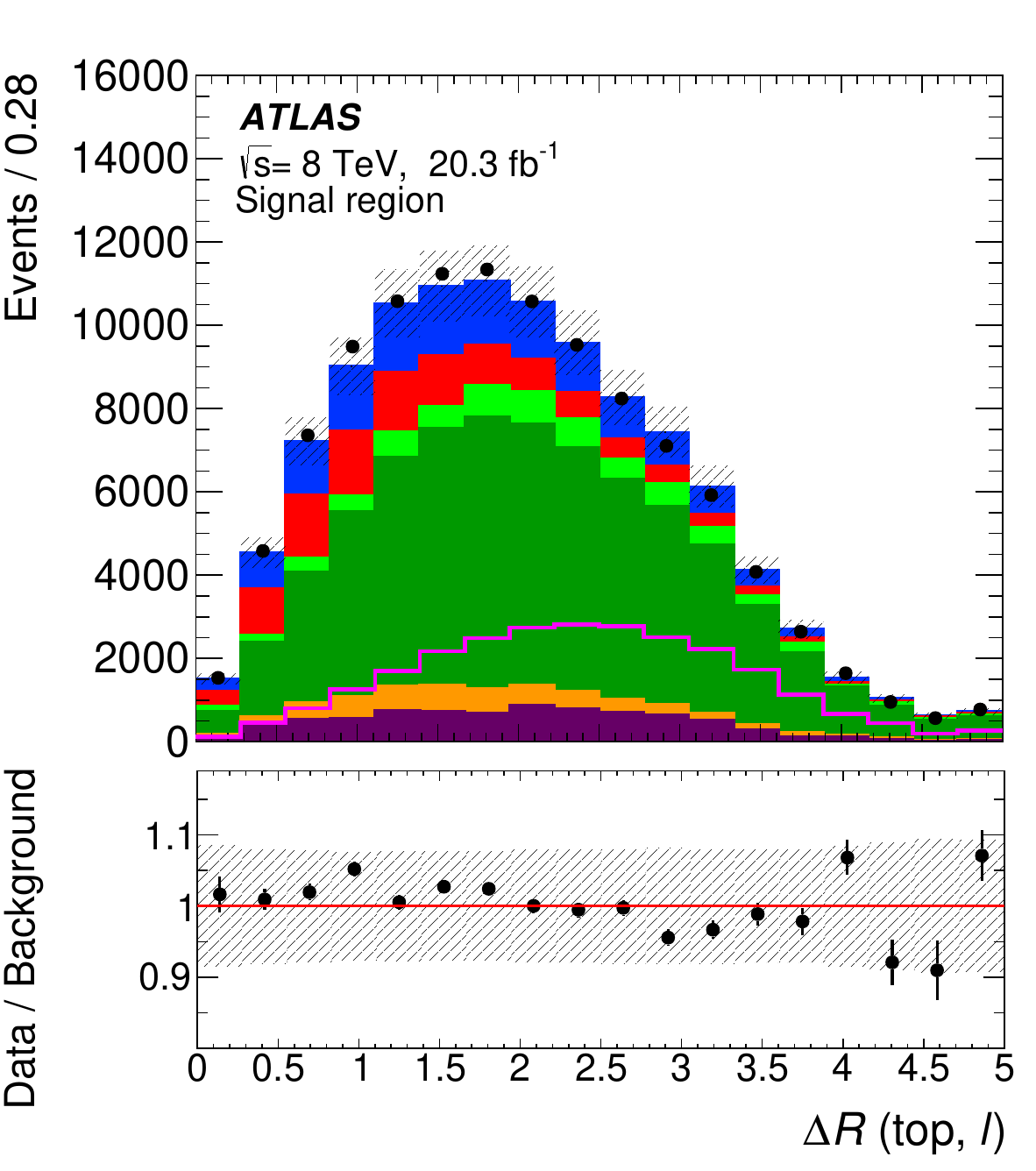}
  }
  \subfigure[][]{\label{subfig:SR_dphijettop}%
    \includegraphics[width=0.32\textwidth]{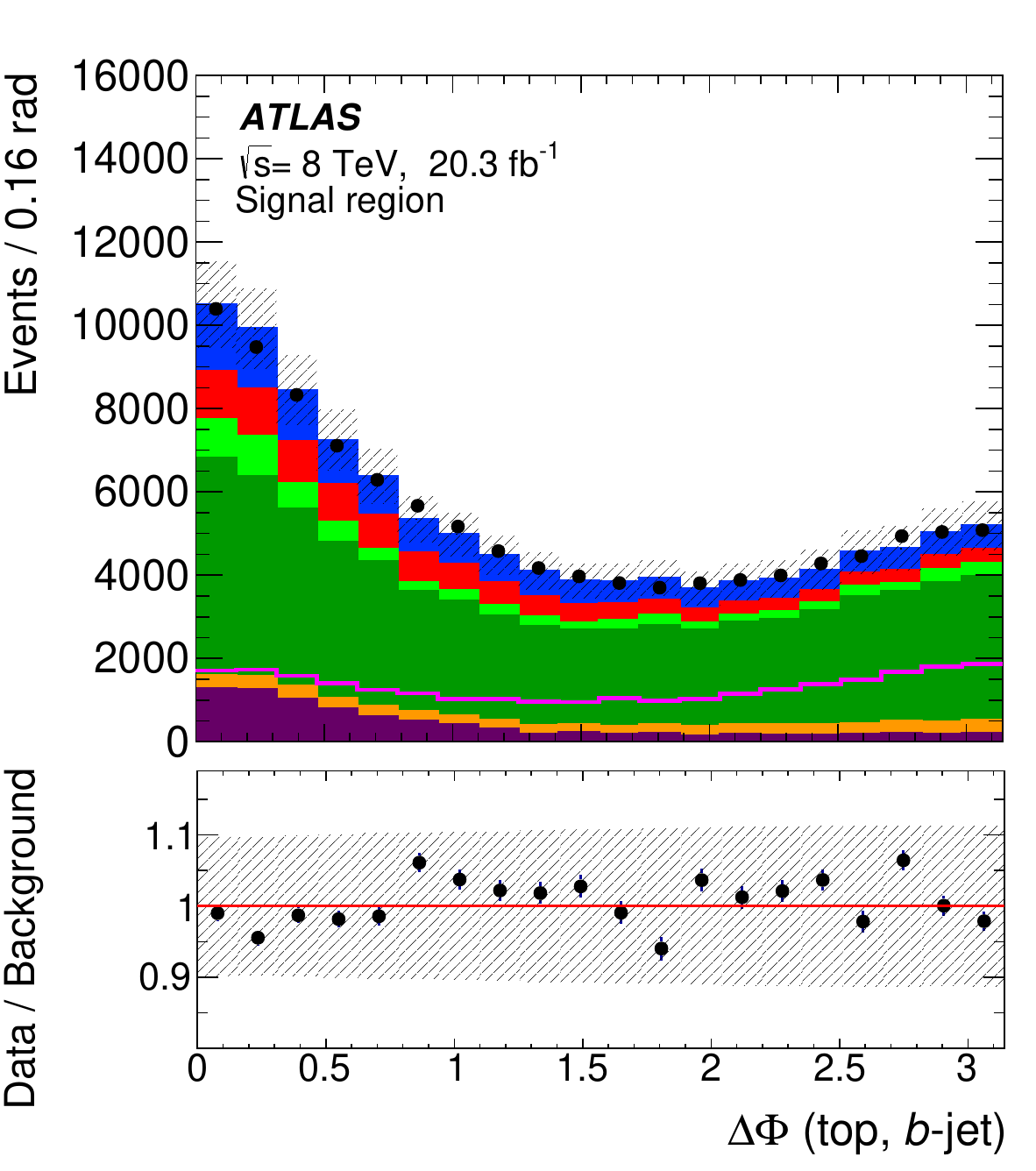}
  }
  \caption{Distributions of three important discriminating variables (except for the transverse momentum of the lepton):
    \subref{subfig:CR_mttop} and \subref{subfig:SR_mttop} the top-quark transverse mass in the control and signal
    regions; \subref{subfig:CR_drleptop} and \subref{subfig:SR_drleptop} the $\Delta R$ between the lepton and the reconstructed top quark in the control and signal regions;
     \subref{subfig:CR_dphijettop} and \subref{subfig:SR_dphijettop} the $\Delta \phi$ between the jet and the
    reconstructed top quark. All processes are normalised using the scale factors obtained in the binned maximum-likelihood fit to the \MET~distribution. The FCNC signal cross-section is scaled to \SI{50}{\pb} and overlayed on the distributions in the signal region.
    The last histogram bin includes overflow events and the hatched band indicates the combined statistical and systematic uncertainties, evaluated after the fit discussed in Section \ref{sec:result}.}
  \label{fig:inputvars}
\end{figure}

The distributions for three of the four most important variables in the control and signal regions 
are shown in \Fig{\ref{fig:inputvars}}. The shape of the multi-jet background is obtained using the 
samples described in Sec.~\ref{sec:bgestimation}.
The distribution of $\pT^{\ell}$ is shown
in \Fig{\ref{fig:control}\subref{subfig:CR_leppt}} for the control region.
The distributions are normalised using the scale factors obtained 
in the binned maximum-likelihood fit to the \MET~distribution.

\begin{table}[htbp]
  \centering
  \begin{tabular}{lp{0.8\textwidth}}
      \toprule
      Variable                              & Definition \\
      \midrule
      $\MT(\text{top})$                     & Transverse mass of the reconstructed top quark \\
      $\pT^{\ell}$                          & Transverse momentum of the charged lepton   \\
      $\Delta R(\text{top},\ell)$           & Distance in the $\eta$--$\phi$ plane between
                                              the reconstructed top quark and the charged lepton\\
      $\pT^{b\text{-jet}}$                  & Transverse momentum of the $b$-tagged jet \\
      $\Delta\phi(\text{top},b\text{-jet})$ & Difference in azimuth between the reconstructed top quark and the $b$-tagged jet\\
      $\cos\theta(\ell,b\text{-jet})$       & Opening angle of the three-vectors between the charged lepton and
                                              the $b$-tagged jet \\
      $q^{\ell}$                            & Charge of the lepton \\
      \MTW                                  & $W$-boson transverse mass \\
      $\eta^{\ell}$                         & Pseudorapidity of the charged lepton \\
      $\Delta\phi(\text{top},W)$            & Difference in azimuth between the reconstructed top quark and the $W$ boson \\
      $\Delta R(\text{top},b\text{-jet})$   & Distance in the $\eta$--$\phi$ plane between
                                              the reconstructed top quark and the $b$-tagged jet\\
      $\eta^{\text{top}}$                   & Pseudorapidity of the reconstructed top quark\\
      $\pT^{W}$                             & Transverse momentum of the $W$ boson \\
  \bottomrule
  \end{tabular}
  \caption{ Variables used in the training of the neural network ordered by their descending importance.}
  \label{tab:trainingVars}
\end{table}

\begin{figure}[htbp]
  \centering
  \includegraphics[width=0.4\textwidth]{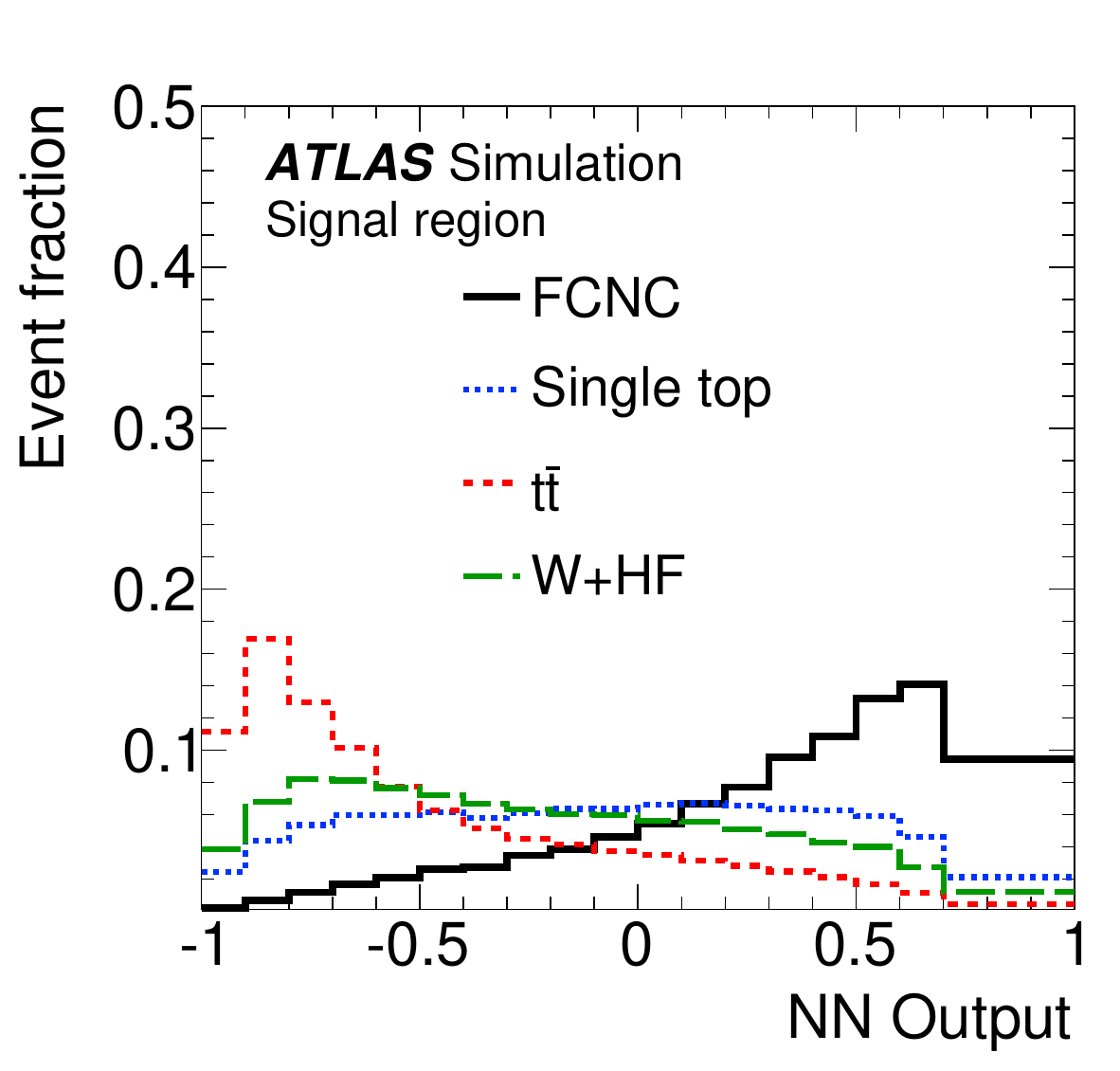}
  \caption{Probability density of the neural-network output distribution for
    the signal and the most important background processes.}
  \label{fig:finalnnshape}
\end{figure}

\begin{figure}[htbp]
  \centering
  \subfigure[][]{\label{subfig:CR_nn}%
    \includegraphics[width=0.45\textwidth]{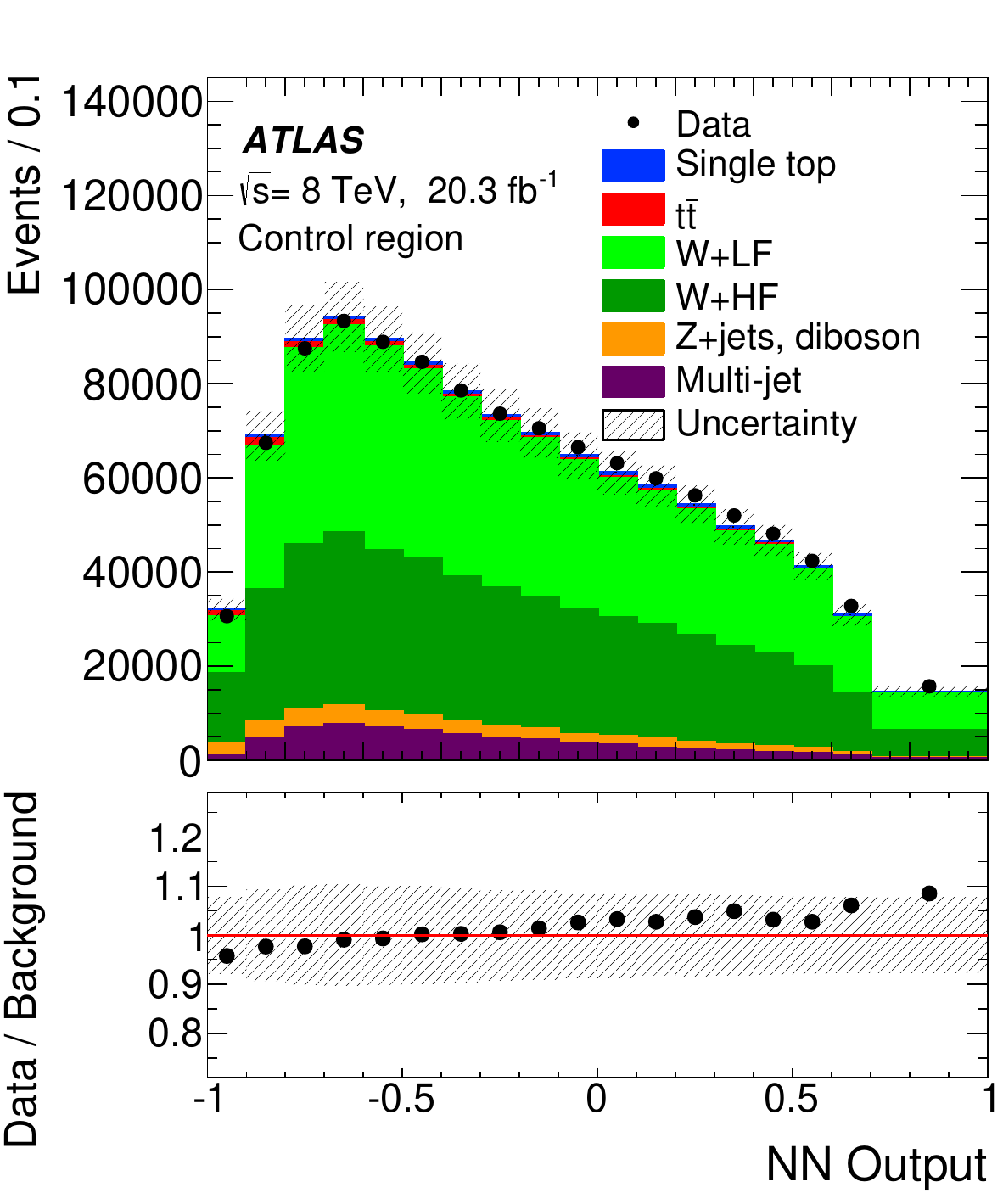}
  }
  \subfigure[][]{\label{subfig:SR_nn}%
    \includegraphics[width=0.45\textwidth]{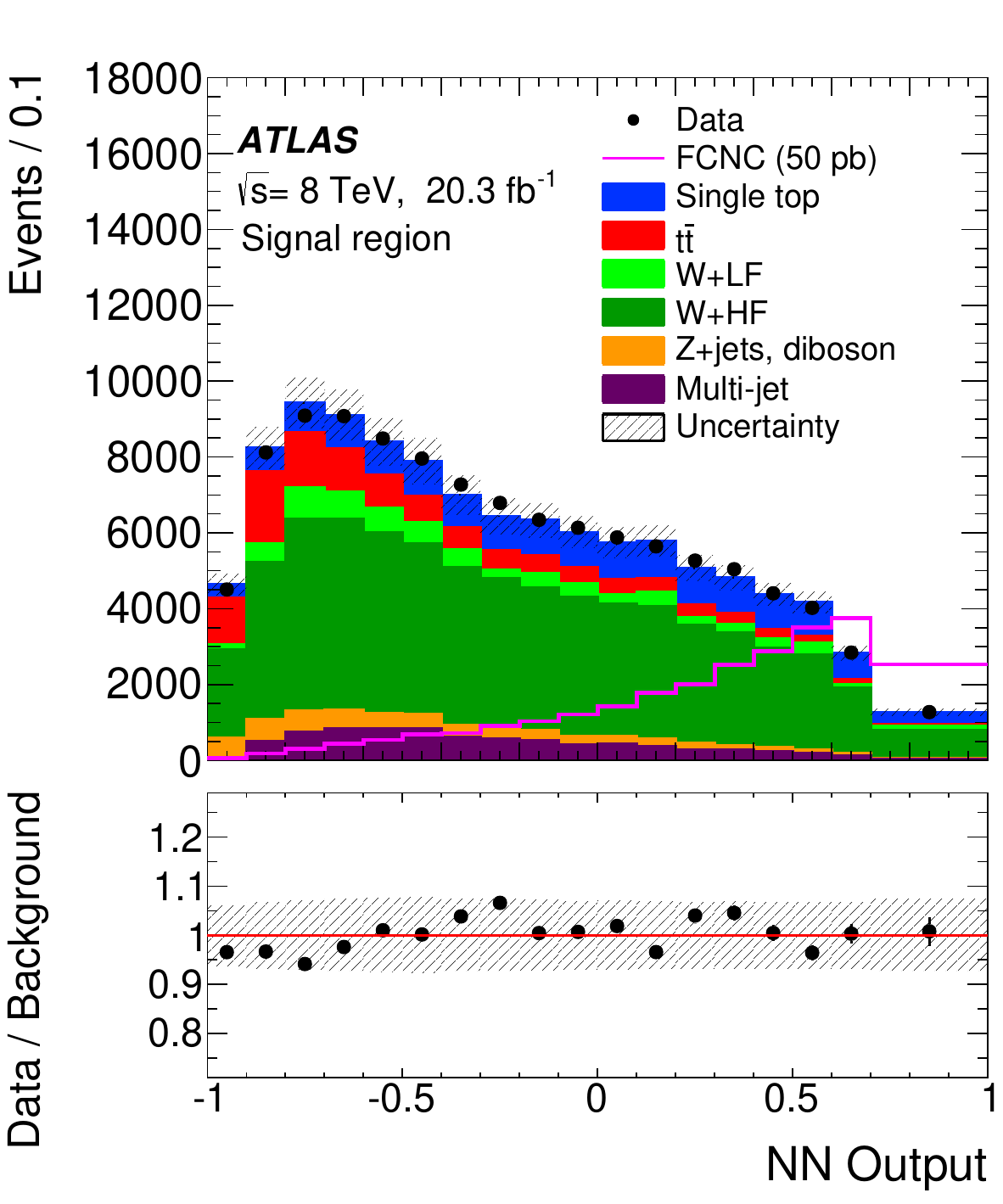}
  }
  \caption{Neural-network output distribution \subref{subfig:CR_nn} in the
    control region and \subref{subfig:SR_nn} in the signal region.
    The shape of the signal scaled to \SI{50}{\pb} is shown in \subref{subfig:SR_nn}.
    All background processes are shown normalised to the result of the binned maximum-likelihood fit used to determine
    the fraction of multi-jet events.
    The hatched band indicates the combined statistical and systematic uncertainties, evaluated after the fit discussed in Section \ref{sec:result}.}
  \label{fig:finalnn}
\end{figure}

The resulting neural-network output distributions for the most important background processes and the signal are
displayed in \Fig{\ref{fig:finalnnshape}} as probability densities and in 
\Figs{\ref{fig:finalnn}\subref{subfig:CR_nn}}{\ref{fig:finalnn}\subref{subfig:SR_nn}} normalised to the number of expected events
in the control and signal regions, respectively.
Signal-like events have output values close to 1, whereas background-like events accumulate near $-1$.  
Overall, good agreement within systematic uncertainties between data and the background processes is observed in
both the control and signal regions.

\section{Systematic uncertainties}
\label{sec:systematics} 

Systematic uncertainties 
are assigned to account for detector calibration and resolution uncertainties, as well as the  
uncertainties on theoretical predictions.
These can affect the normalisation of the individual
backgrounds and the signal acceptance (acceptance uncertainties) as well as the shape of the neural-network output 
distribution (shape uncertainties).
Quoted relative uncertainties refer to acceptance of the respective processes unless stated otherwise. 

\subsection{Object modelling}

The effects of the systematic uncertainties due to the residual differences between 
data and Monte Carlo simulation, uncertainties on jets, electron and muon 
reconstruction after calibration, and uncertainties on scale 
factors that are applied to the simulation are estimated using pseudo-experiments.

Uncertainties on the muon (electron) trigger, reconstruction and 
selection efficiency scale factors are estimated in measurements of
$Z \rightarrow \mu\mu$ ($Z \rightarrow e e$ and $W \rightarrow e\nu$) production.
The scale factor uncertainties are as large as \SI{5}{\%}.
To evaluate uncertainties on the lepton momentum scale and resolution,
the same processes are used~\cite{PERF-2013-05}. The uncertainty on the 
charge misidentification acceptances were studied and found to be negligible
for this analysis. 

The jet energy scale (JES) is derived
using information from test-beam data, LHC collision data and
simulation. Its uncertainty varies between \SI{2.5}{\%} and \SI{8}{\%},
depending on jet \pT and $\eta$~\cite{PERF-2012-01}. This includes uncertainties in 
the fraction of jets induced by gluons and mismeasurements due to 
close-by jets. Additional uncertainties due to pile-up can be as
large as \SI{5}{\%}. 
An additional jet energy scale uncertainty of up to \SI{2.5}{\%}, depending on the \pT of the jet,
is applied for $b$-quark-induced jets due to differences between light-quark
and gluon jets compared to jets containing $b$-hadrons.
Additional uncertainties are from the modelling of the jet energy resolution and the
missing transverse momentum, which accounts for contributions of calorimeter cells 
not matched to any jets, soft jets, and pile-up.
The effect of uncertainties associated with the jet-vertex fraction is also considered for each jet.

Since the analysis makes use of $b$-tagging, the uncertainties on the $b$- and $c$-tagging 
efficiencies and the mistag acceptance~\cite{ATLAS-CONF-2014-046,ATLAS-CONF-2014-004} 
are taken into account.  

\subsection{Multi-jet background}

For the multi-jet background, an uncertainty on the estimated multi-jet
fractions and the modelling is included.
The systematic uncertainty on the fractions,
as well as a shape uncertainty, are obtained by comparing to an alternative method, 
the matrix method~\cite{ATLAS-CONF-2014-058}.
The method estimates the number of multi-jet background events in the signal region based on loose and tight lepton
isolation definitions, the latter selection being a subset of the former.
The number of multi-jet events $N^\text{tight}_\text{fake}$ passing the tight (signal) isolation requirements can be expressed as:
\begin{linenomath}
\begin{equation*}
N^\text{tight}_\text{fake} = \frac{\epsilon_\text{fake}}{\epsilon_\text{real} - \epsilon_\text{fake}} \cdot (N^\text{loose} \epsilon_\text{real} - N^\text{tight})\,,
\end{equation*}
\end{linenomath}
where $\epsilon_\text{real}$ and $\epsilon_\text{fake}$ are the efficiencies for real and fake loose leptons being
selected as tight leptons, $N^\text{loose}$ is the number of selected events in the loose sample, and $N^\text{tight}$ is the number of selected events in the signal sample.
By comparing the two methods, the uncertainty on the fraction of multi-jet events is estimated to be \SI{17}{\%}.
The shape uncertainty is constructed by comparing the neural-network
output distributions of the jet-lepton and anti-muon samples with the distributions 
obtained using the matrix method.

\subsection{Monte Carlo generators}

Systematic effects from the modelling of the signal and background processes
are taken into account by comparing different generator models and varying the 
parameters of the event generation.
The effect of parton-shower modelling for the 
top-quark processes is tested by comparing two \POWHEG samples interfaced
to \HERWIG and \PYTHIA, respectively. There are also differences associated with the way in which double-counted events in the NLO corrections and the parton showers are removed.
These are estimated by comparing samples produced with the \MCatNLO method and the \POWHEG method.

The difference between the top-quark mass used in the simulations and the measured value has negligible effect on the results.

For the single top-quark processes, variations of initial- and final-state radiation (ISR and FSR)
together with variations of the hard-process scale are studied.
The uncertainty is estimated using events generated with \POWHEG interfaced to \PYTHIA. 
Factorisation and renormalisation scales are varied independently by factors of $0.5$ and $2.0$, 
while the scale of the parton shower is varied consistently with the 
renormalisation scale using specialised Perugia 2012 tunes~\cite{Skands:2010ak}.
The uncertainty on the amounts of ISR and FSR in the simulated \ttbar sample is assessed using
\ALPGEN samples, showered  with \PYTHIA,  with  varied  amounts  of  initial-  and
final-state radiation, which are compatible with the measurements of additional jet activity
in \ttbar events~\cite{TOPQ-2011-21}.

The effect of applying the $W$-boson \pT reweighting was studied and found 
to have negligible impact on the shape of the neural-network output distribution and the measured cross-section.
Hence no systematic uncertainty due to this was assigned.
 
Finally, an uncertainty is included to account for statistical effects from the limited size of the MC samples.

\subsection{Parton distribution functions}

Systematic uncertainties related to the parton distribution functions
are taken into account for all samples using simulated events. The events are
reweighted according to each of the PDF uncertainty eigenvectors or replicas and the
uncertainty is calculated following
the recommendation of the respective PDF group~\cite{Botje:2011sn}.
The final PDF uncertainty is given by the envelope of the estimated uncertainties for the \ct10 PDF set, 
the \mstw PDF set and the \nnpdf PDF set.

\subsection{Theoretical cross-section normalisation}

The theoretical cross-sections and their uncertainties are given in \Sect{\ref{sec:bgestimation}} for each background 
process.
Since the single top-quark $t$-, $Wt$-, and $s$-channel processes are grouped together in the statistical analysis, 
their uncertainties are added in proportion to their relative fractions, leading to a combined uncertainty
of \SI{10}{\%}. 

A cross-section uncertainty of \SI{4}{\%} is assigned for the $W/Z$+(0 jet) process, 
while \textsc{ALPGEN} parameter variations of the 
factorisation and renormalisation scale and the matching parameter consistent with experimental data 
yield an uncertainty on the cross-section ratio of \SI{24}{\%}.
For \WHF production, a conservatively estimated uncertainty on the HF fraction of \SI{50}{\%} is added.
This uncertainty is also applied to the combined $Z$+jets and diboson background.

\subsection{Luminosity}

The uncertainty on the measured luminosity is estimated to be \SI{2.8}{\%}.
It is derived from beam-separation scans performed in November 2012, following the same methodology as that detailed in Ref.~\cite{DAPR-2011-01}.



\section{Results}
\label{sec:result}

In order to estimate the signal content of the selected sample, a binned maximum-likelihood fit
to the complete neural-network output distributions in the signal region is performed.
Including all bins of the neural-network output distributions 
in the fit has the advantage of making maximal use of all signal events remaining after the event selection,
and, in addition, allows the background acceptances to be constrained by the data.

The signal rates, the rate of the single top-quark and \ttbar background and the rate of the \WHF background are fitted simultaneously. 
The event yields of the multi-jet background, the \WLF and the combined $Z$+jets/diboson 
background are not allowed to vary in the fit, but instead are fixed to the estimates 
given in Table~\ref{tab:evtyield}.

No significant rate of FCNC single top-quark production is observed.
An upper limit is set using hypothesis tests.
The compatibility of the data with the signal hypothesis, which depends
on the coupling constants, and the background hypothesis is 
evaluated by performing a frequentist hypothesis test based
on pseudo-experiments,
corresponding to an integrated luminosity of \SI{20.3}{\per\fb}.
Two hypotheses are compared: the null hypothesis, $H_0$, and
the signal hypothesis, $H_1$, which includes FCNC single top-quark production.
For both scenarios, ensemble tests, i.e.\ large sets of pseudo-experiments, are performed.
Systematic uncertainties are included in the pseudo-experiments using variations of the
signal acceptance, the background acceptances and the shape of the neural-network output distribution
due to all sources of uncertainty described in the previous section.

To distinguish between the two hypotheses, the so-called $Q$ value is used as a test statistic.
It is defined as the ratio of the likelihood function $L$, evaluated for the different hypotheses:
\begin{linenomath}
\begin{equation}
  Q = -2 \ln \left(\frac{L\left( \beta^\text{FCNC} = 1 \right)}{ L\left( \beta^\text{FCNC} = 0 \right)} \right),
\end{equation}
\end{linenomath}
where $\beta^\text{FCNC}$ is the scale factor for 
the number of events expected from the signal process
for an assumed production cross-section.
Systematic uncertainties are included by varying the predicted 
number of events for the signal and all background processes in the pseudo-experiments. 

The \CLs method~\cite{cls} is used to derive confidence
levels (\CL) for a certain value of $Q^{\text{obs}}$ and $Q^{\text{exp}}$.
A particular signal hypothesis $H_1$, determined by given coupling constants $\kappa_{ugt}/\Lambda$ and
$\kappa_{cgt}/\Lambda$, is excluded at the \SI{95}{\%} \CL if a $\CLs < 0.05$ is found. The observed \SI{95}{\%} \CL upper limit on the anomalous FCNC single top-quark production cross-section multiplied by the  $t \rightarrow  \ell\nu b$ branching fraction,
including all uncertainties, is \SI{3.4}{\pb}, while the expected upper limit is
\SI[parse-numbers=false]{2.9^{+1.9}_{-1.2}}{\pb}.

\begin{figure}[htbp]
  \centering
  \subfigure[][]{\label{subfig:NNlimitcontrol}%
    \includegraphics[width=0.45\textwidth]{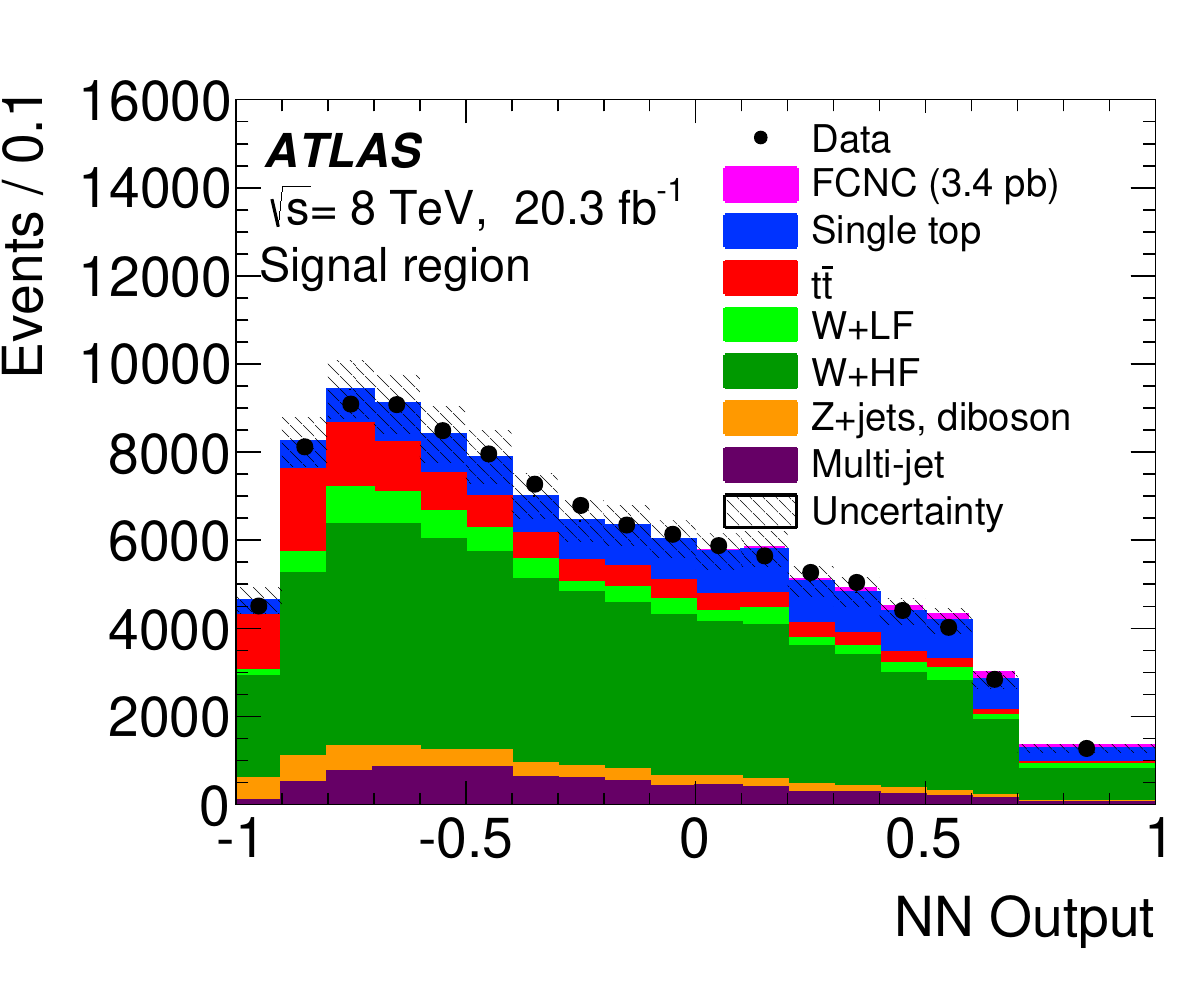}
  }
  \subfigure[][]{\label{subfig:NNlimitsignal}%
  \includegraphics[width=0.45\textwidth]{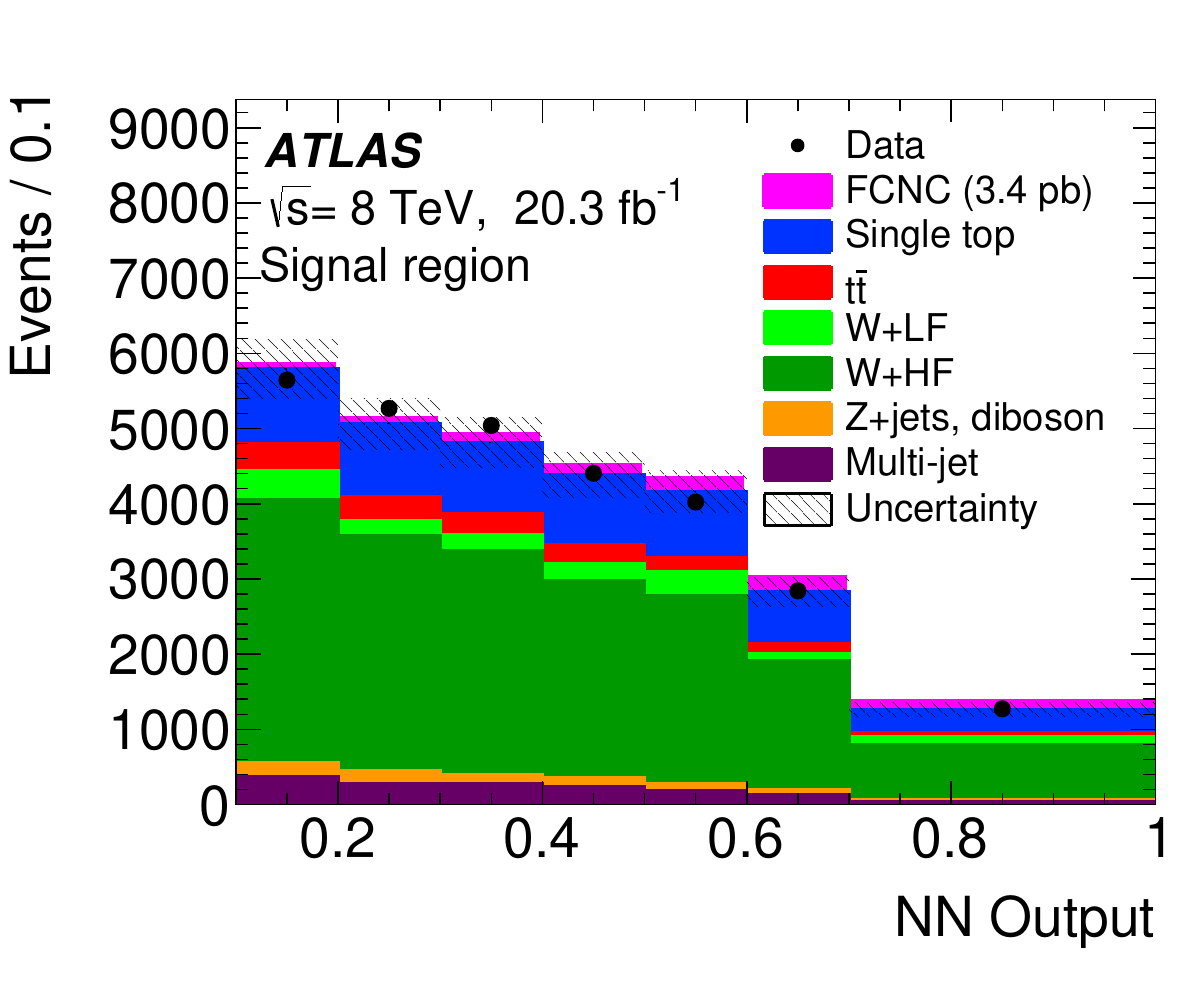}
  }
  \caption{\subref{subfig:NNlimitcontrol} Neural-network output distribution in the signal region 
  and \subref{subfig:NNlimitsignal} in the signal region with neural network output above 0.1.
  In both figures the signal contribution scaled to the observed upper limit is shown.
  The hatched band indicates the total posterior uncertainty as obtained from the limit calculation.}
  \label{fig:finalnnlimitscales}
\end{figure}

To visualise the observed upper limit in the neural-network output distribution, the FCNC signal process scaled to
\SI{3.4}{\pb} stacked on top of all background processes is shown in \Fig{\ref{fig:finalnnlimitscales}}. 

The total uncertainty is dominated by the jet energy resolution uncertainty, the modelling of $\MET$
and the uncertainty on the normalisation and the modelling of the multi-jet background.
A summary of all considered sources and their impact on the expected upper limit is
shown in \Tab{\ref{tab:limitunc}}.

\begin{table}[htbp]
  \centering
  \sisetup{round-mode=figures}
\begin{tabular}{lS[round-precision=2]S[table-format=2.0,round-precision=0,round-mode=places]}
	\toprule
	Source                                 & {Expected \SI{95}{\%} \CL upper limit} & {Change in the upper limit} \\
	                                       & [\si{\pb}]                             & {[\%]}                      \\
	\midrule
	Normalisation \& MC statistics         & 1.5                                    & {-}                         \\
	\midrule
	Multi-jets normalisation and modelling & 1.8                                    & 25.3                        \\
	                             \multicolumn{2}{c}{ }                              &  \\
	Luminosity                             & 1.5                                    & 4.79                        \\
	                             \multicolumn{2}{c}{ }                              &  \\
	Lepton identification                  & 1.5                                    & 3.42                        \\
	Electron energy scale                  & 1.6                                    & 8.22                        \\
	Electron energy resolution             & 1.5                                    & 4.12                        \\
	Muon momentum scale                    & 1.5                                    & 1.37                        \\
	Muon momentum resolution               & 1.5                                    & 4.79                        \\
	                             \multicolumn{2}{c}{ }                              &  \\
	Jet energy scale                       & 1.6                                    & 7.53                        \\
	Jet energy resolution                  & 1.9                                    & 32.2                        \\
	Jet reconstruction efficiency          & 1.5                                    & 4.12                        \\
	Jet vertex fraction scale              & 1.5                                    & 3.42                        \\
	                             \multicolumn{2}{c}{ }                              &  \\
	$b$-tagging efficiency                 & 1.5                                    & 3.42                        \\
	$c$-tagging efficiency                 & 1.5                                    & 4.12                        \\
	Mistag acceptance                      & 1.5                                    & 2.05                        \\
	                             \multicolumn{2}{c}{ }                              &  \\
	$\MET$ modelling                       & 1.9                                    & 33.6                        \\
	                             \multicolumn{2}{c}{ }                              &  \\
	PDF                                    & 1.5                                    & 5.48                        \\
	Scale variations                       & 1.5                                    & 2.05                        \\
	MC generator (NLO subtraction method)       & 1.6                                    & 7.53                        \\
	Parton shower modelling                & 1.5                                    & 4.79                        \\
	\midrule
	All systematic uncertainties           & 2.9                                    & {-}                         \\
	\bottomrule
\end{tabular}
  \caption{The effect of a single systematic uncertainty in addition to the cross-section normalisation
           and MC statistical uncertainties alone (top row) on the expected \SI{95}{\%} \CL 
           upper limits on the anomalous FCNC single top-quark production $qg\to t \to b\ell\nu$. The relative change quoted in the third column is with respect to the expected limit with normalisation and MC statistical uncertainties only.}
  \label{tab:limitunc}
\end{table}

\begin{figure}[tbp]
  \centering
  \subfigure[][]{\label{subfig:coupling}%
    \includegraphics[width=0.45\textwidth]{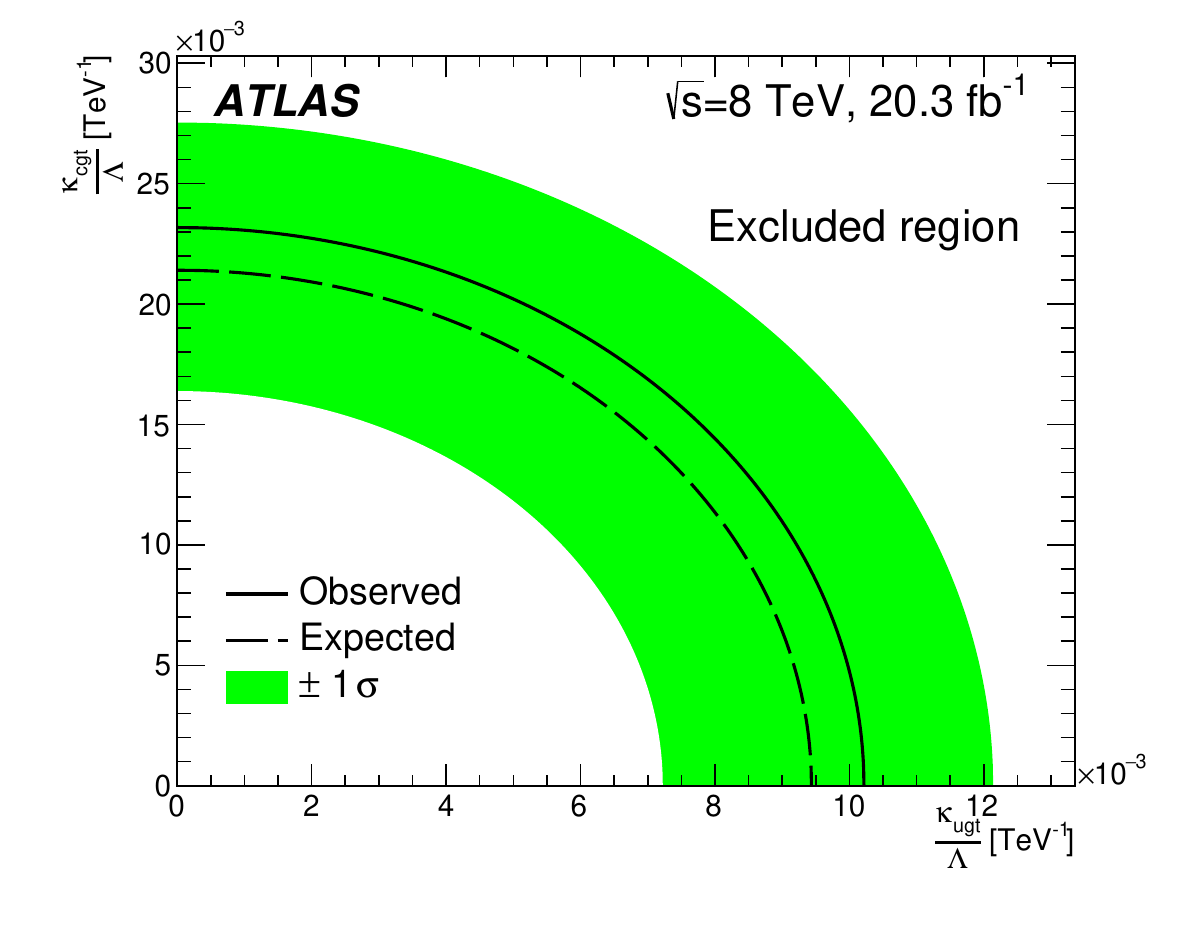}
  }
  \subfigure[][]{\label{subfig:branching}%
   \includegraphics[width=0.45\textwidth]{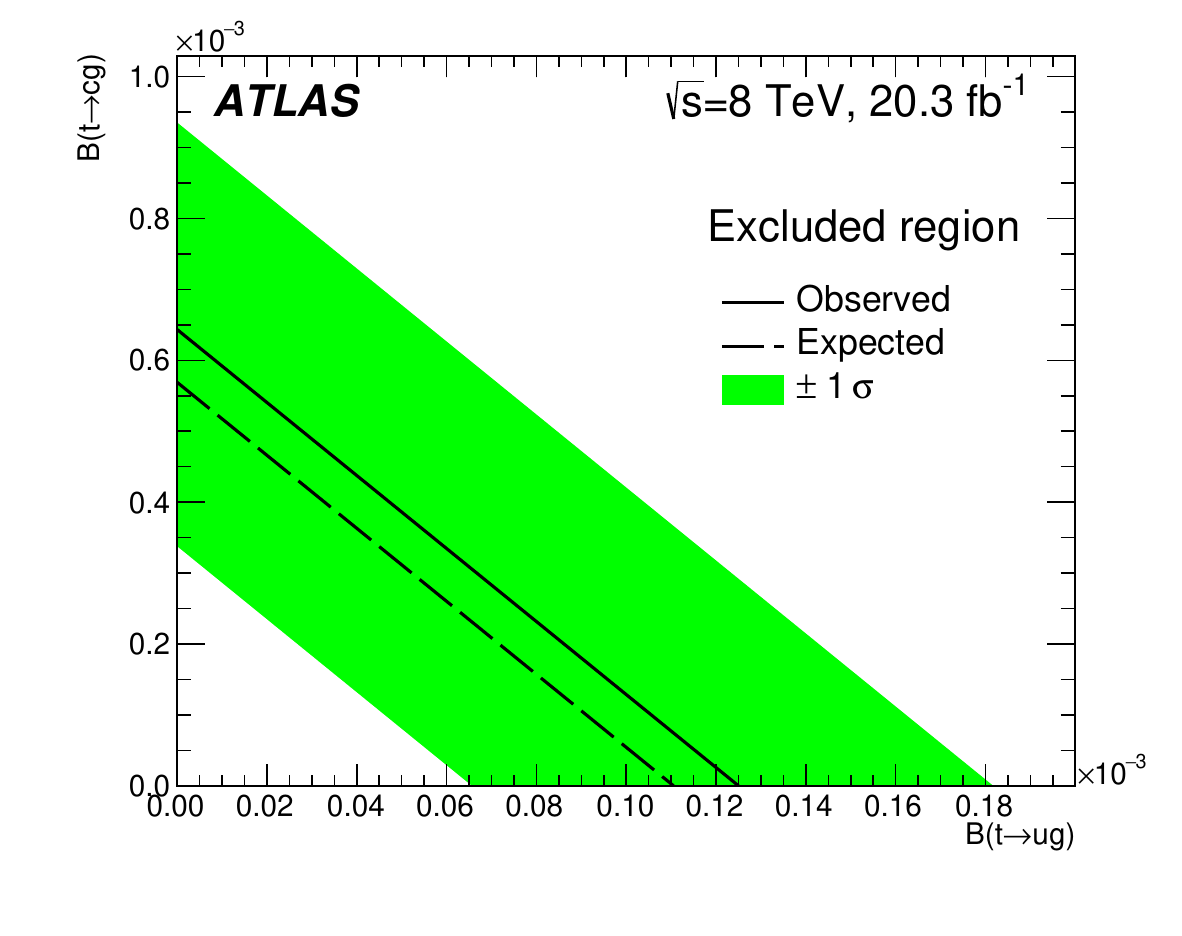}
  }
  \caption{\subref{subfig:coupling} Upper limit on the coupling constants $\kappa_{ugt}$ and $\kappa_{cgt}$ and 
  \subref{subfig:branching} on the branching fractions $\BR(t \rightarrow ug)$ and $\BR(t \rightarrow cg)$.
  The shaded band shows the one standard deviation variation of the expected limit.}
  \label{fig:final_couplings}
\end{figure}

Using the NLO predictions for the FCNC single top-quark production cross-section~\cite{Gao:2011fx,Liu:2005dp} and assuming 
$\BR(t \to Wb) = 1$, the upper limit on the cross-section can be interpreted as a limit 
on the coupling constants divided by the scale of new physics:
$\kappa_{ugt}/\Lambda < \SI{10E-3}{\per\TeV}$ assuming $\kappa_{cgt}/\Lambda = 0$,
and $\kappa_{cgt}/\Lambda < \SI{23E-3}{\per\TeV}$ assuming $\kappa_{ugt}/\Lambda = 0$.
Distributions of the upper limits on the coupling constants for combinations of $cgt$ and $ugt$ channels are shown in \Fig{\ref{fig:final_couplings}\subref{subfig:coupling}}.

Limits on the coupling constants can also be interpreted as limits on the branching fractions
using $\BR(t \rightarrow qg) = \mathcal{C} \left(\kappa_{qgt} / \Lambda\right)^{2}$, where
$\mathcal{C}$ is calculated at NLO~\cite{Zhang:2008yn}.
Upper limits on the branching fractions
$\BR(t \rightarrow ug) < \num{1.2E-4}$, assuming $\BR(t \rightarrow cg)=0$ and 
$\BR(t \rightarrow cg) < \num{6.4E-4}$, assuming $\BR(t \rightarrow ug)=0$, are derived
and presented in \Fig{\ref{fig:final_couplings}\subref{subfig:branching}}.

\FloatBarrier

\section{Conclusion}
\label{sec:conclusion}

A search for anomalous single top-quark production via strong flavour-changing neutral currents 
in $pp$ collisions at the LHC is performed.
Data collected by the ATLAS experiment in 2012 at a centre-of-mass energy $\sqrt{s} =
\SI{8}{\TeV}$, and corresponding to an integrated luminosity of \SI{20.3}{\per\fb} are used.
Candidate events for which a $u$- or $c$-quark interacts with a gluon to produce a single top quark are selected.
To discriminate between signal and background processes, a multivariate technique using a neural network is applied.
The final statistical analysis is performed using a frequentist technique.
As no signal is seen in the observed output distribution, an upper limit on the production cross-section is
set. The expected  \SI{95}{\%} \CL limit on the production cross-section multiplied by the $t \to b \ell \nu$ branching fraction 
is $\sigma_{qg \to t} \times \BR(t \to bW) \times \BR(W \to \ell\nu) < \SI{2.9}{\pb}$ and the observed \SI{95}{\%} \CL limit is 
$\sigma_{qg \rightarrow t} \times \BR (t \rightarrow bW) \times \BR(W \to \ell\nu) < \SI{3.4}{\pb}$.
Upper limits on the coupling constants divided by the scale of new physics
$\kappa_{ugt}/\Lambda < \SI{10e-3}{\per\TeV}$ and
$\kappa_{cgt}/\Lambda < \SI{23e-3}{\per\TeV}$
and on the branching fractions
$\BR(t \rightarrow ug) < \num{1.2E-4} $ and
$\BR(t \rightarrow cg) < \num{6.4E-4} $ 
are derived from the observed limit.
These are the most stringent limits published to date.

\section*{Acknowledgements}


We thank CERN for the very successful operation of the LHC, as well as the
support staff from our institutions without whom ATLAS could not be
operated efficiently.

We acknowledge the support of ANPCyT, Argentina; YerPhI, Armenia; ARC, Australia; BMWFW and FWF, Austria; ANAS, Azerbaijan; SSTC, Belarus; CNPq and FAPESP, Brazil; NSERC, NRC and CFI, Canada; CERN; CONICYT, Chile; CAS, MOST and NSFC, China; COLCIENCIAS, Colombia; MSMT CR, MPO CR and VSC CR, Czech Republic; DNRF, DNSRC and Lundbeck Foundation, Denmark; IN2P3-CNRS, CEA-DSM/IRFU, France; GNSF, Georgia; BMBF, HGF, and MPG, Germany; GSRT, Greece; RGC, Hong Kong SAR, China; ISF, I-CORE and Benoziyo Center, Israel; INFN, Italy; MEXT and JSPS, Japan; CNRST, Morocco; FOM and NWO, Netherlands; RCN, Norway; MNiSW and NCN, Poland; FCT, Portugal; MNE/IFA, Romania; MES of Russia and NRC KI, Russian Federation; JINR; MESTD, Serbia; MSSR, Slovakia; ARRS and MIZ\v{S}, Slovenia; DST/NRF, South Africa; MINECO, Spain; SRC and Wallenberg Foundation, Sweden; SERI, SNSF and Cantons of Bern and Geneva, Switzerland; MOST, Taiwan; TAEK, Turkey; STFC, United Kingdom; DOE and NSF, United States of America. In addition, individual groups and members have received support from BCKDF, the Canada Council, CANARIE, CRC, Compute Canada, FQRNT, and the Ontario Innovation Trust, Canada; EPLANET, ERC, FP7, Horizon 2020 and Marie Skłodowska-Curie Actions, European Union; Investissements d'Avenir Labex and Idex, ANR, Region Auvergne and Fondation Partager le Savoir, France; DFG and AvH Foundation, Germany; Herakleitos, Thales and Aristeia programmes co-financed by EU-ESF and the Greek NSRF; BSF, GIF and Minerva, Israel; BRF, Norway; the Royal Society and Leverhulme Trust, United Kingdom.

The crucial computing support from all WLCG partners is acknowledged
gratefully, in particular from CERN and the ATLAS Tier-1 facilities at
TRIUMF (Canada), NDGF (Denmark, Norway, Sweden), CC-IN2P3 (France),
KIT/GridKA (Germany), INFN-CNAF (Italy), NL-T1 (Netherlands), PIC (Spain),
ASGC (Taiwan), RAL (UK) and BNL (USA) and in the Tier-2 facilities
worldwide.

\printbibliography

\clearpage
\begin{flushleft}
{\Large The ATLAS Collaboration}

\bigskip

G.~Aad$^{\textrm 85}$,
B.~Abbott$^{\textrm 113}$,
J.~Abdallah$^{\textrm 151}$,
O.~Abdinov$^{\textrm 11}$,
R.~Aben$^{\textrm 107}$,
M.~Abolins$^{\textrm 90}$,
O.S.~AbouZeid$^{\textrm 158}$,
H.~Abramowicz$^{\textrm 153}$,
H.~Abreu$^{\textrm 152}$,
R.~Abreu$^{\textrm 116}$
% ...
\bigskip
\\
% ...
$^{11}$ Institute of Physics, Azerbaijan Academy of Sciences, Baku, Azerbaijan\\
% ...
$^{85}$ CPPM, Aix-Marseille Universit{\'e} and CNRS/IN2P3, Marseille, France\\
% ...
$^{90}$ Department of Physics and Astronomy, Michigan State University, East Lansing MI, United States of America\\
% ...
$^{107}$ Nikhef National Institute for Subatomic Physics and University of Amsterdam, Amsterdam, Netherlands\\
% ...
$^{113}$ Homer L. Dodge Department of Physics and Astronomy, University of Oklahoma, Norman OK, United States of America\\
% ...
$^{116}$ Center for High Energy Physics, University of Oregon, Eugene OR, United States of America\\
% ...
$^{151}$ Institute of Physics, Academia Sinica, Taipei, Taiwan\\
$^{152}$ Department of Physics, Technion: Israel Institute of Technology, Haifa, Israel\\
$^{153}$ Raymond and Beverly Sackler School of Physics and Astronomy, Tel Aviv University, Tel Aviv, Israel\\
% ...
$^{158}$ Department of Physics, University of Toronto, Toronto ON, Canada\\
% ...
\end{flushleft}


\end{document}